\documentclass[10pt,a4paper]{article}
\usepackage{amssymb,amsfonts}
\begin{document}
\title{Localized Endomorphisms of the Chiral Ising Model}
\author{Jens B{\"o}ckenhauer\\II. Institut f{\"u}r
Theoretische Physik, Universit{\"a}t Hamburg\\
Luruper Chaussee 149, D-22761 Hamburg}
\maketitle

\begin{abstract}
Based on the treatment of the chiral Ising model
by Mack and Schomerus, we present examples of localized
endomorphisms $\varrho_1^{\rm loc}$ and
$\varrho_{1/2}^{\rm loc}$. It is shown that they
lead to the same superselection sectors as the global
ones in the sense that unitary equivalence
$\pi_0\circ\varrho_1^{\rm loc}\cong\pi_1$ and
$\pi_0\circ\varrho_{1/2}^{\rm loc}\cong\pi_{1/2}$ holds.
Araki's formalism of the selfdual CAR algebra is used
for the proof. We prove local normality and extend
representations and localized endomorphisms to a
global algebra of observables which
is generated by local von Neumann algebras on the
punctured circle. In this framework, we manifestly
prove fusion rules and derive statistics operators.
\end{abstract}

\newcommand{\I}{{\rm i}}
\newcommand{\E}{{\rm e}}
\newtheorem{definition}{Definition}[section]
\newtheorem{theorem}[definition]{Theorem}
\newtheorem{lemma}[definition]{Lemma}
\newtheorem{corollary}[definition]{Corollary}
\newtheorem{proposition}[definition]{Proposition}

\section{Introduction}
In local quantum field theory one considers a Hilbert
space ${\cal H}$ of physical states which decomposes
into orthogonal subspaces ${\cal H}_J$ (superselection
sectors) so that observables do not make transitions
between the sectors. The subspaces ${\cal H}_J$ carry
inequivalent, irreducible representations of the
observable algebra ${\cal A}$, possibly with some
multiplicities \cite{HK}. Among the superselection sectors,
there is a distinguished sector ${\cal H}_0$ which
contains the vacuum vector $|\Omega_0\rangle$ and
carries the vacuum representation $\pi_0$.

The starting point in the algebraic approach to quantum
field theory is the observable algebra ${\cal A}$ which
is usually defined as the $C^*$-inductive limit of the
net of local von Neumann algebras
$\{{\cal A}({\cal O}),{\cal O}\in{\cal K}\}$, where
${\cal K}$ denotes the set of open double cones in $D$
dimensional Minkowski space. The net is assumed to
satisfy the Haag-Kastler-axioms. In general, the observable
algebra ${\cal A}$ admits a lot of inequivalent
representations. Therefore one has to find an appropriate
selection criterion which rules out the physically
non-relevant representations. Doplicher, Haag and Roberts
\cite{DHR1,DHR2,Haag} developed the theory of locally
generated sectors; they suggested that one has to consider
only those representations $\pi_J$ which become equivalent
to the vacuum representation in the restriction to the
causal complement ${\cal O}'$ of any sufficiently large
double cone ${\cal O}\in{\cal K}$. That means that for
a representation $\pi_J$ satisfying the DHR criterion,
there exists for each sufficiently large double cone
${\cal O}$ a unitary $V:{\cal H}_0\rightarrow{\cal H}_J$
such that
\[\pi_J(A)=V\pi_0(A)V^*,\qquad A\in{\cal A}({\cal O}').\]
The DHR criterion leads to the characterization of
superselection sectors by localized endomorphisms:
Usually ${\cal A}$ and $\pi_0({\cal A})$ are identified,
and one defines
\[ \varrho_J(A) = V^* \pi_J(A) V,\qquad A\in{\cal A}.\]
Then $\varrho_J$ is an endomorphism of the observable
algebra and it is localized in ${\cal O}$ in the sense
that $\varrho_J(A)=A$ for all $A\in{\cal A}({\cal O}')$.
Moreover, $\pi_0\circ\varrho_J$ is a representation of
${\cal A}$ in ${\cal H}_0$ which is equivalent to $\pi_J$.
The use of DHR endomorphisms allows to extract all
physical information out of the vacuum sector and to
work without charged, unobservable fields. It is another
advantage that endomorphisms can be composed; it is
possible to formulate fusion rules in terms of equivalence
classes of localized endomorphisms.

Unfortunately, it seems to be very difficult to construct
these endomorphisms explicitly in models. Although the
conformal field theory has turned out to be an interesting
and fruitful area of application of the DHR program,
examples of localized endomorphisms which generate
charged sectors are known explicitly only for a rather
small number of models, e.g.~the $U(1)$ current algebra
on the circle \cite{BMT}. Endomorphisms
have been constructed for Level 1 WZW models \cite{FGV} and,
before that, for the chiral Ising model \cite{MS1,MS2},
however, they are in no sense localized. Mack and Schomerus
had already described the construction of localized
endomorphisms for the chiral Ising model in \cite{MS1},
but it has not yet been proven that they lead
to the same sectors as the global ones and, in particular,
that they lead to irreducible representations. This is
done in the present paper.

In two-dimensional conformal field theory one considers
as basic observable the stress energy tensor which
generates the space time symmetry. Its light
cone components $T_\pm(z_\pm)$ live separately on the
compactified light cone variables $z_\pm\in S^1$, one
deals with chiral fields.
Treating each component for its own, the stress energy
tensor has well-known commutation relations, fixed
up to a constant $c$ \cite{Berti,Mack};
the stress energy tensor generates the Virasoro
algebra {\sf Vir}. In the case $c=\frac{1}{2}$
(Ising model) the Virasoro algebra admits three
inequivalent positive energy representations
$\pi_J$, $J=0,\frac{1}{2},1$, which are lowest weight
representations; $\pi_0$ is identified to be the
vacuum representation. In the chiral Ising model, the
stress energy tensor can be built of a free fermion
field, the Majorana field \cite{MS1,Mack}.
Smearing out the Majorana
field with test functions having support in a
proper subinterval $I\subset S^1$ and considering
bilinear expressions of it, these objects generate
the local $C^*$-algebra ${\cal A}(I)$ of observables.
Such local algebras ${\cal A}(I)$ generate
a global observable algebra ${\cal A}_{\rm univ}^{C^*}$.
Unfortunately, the Virasoro generators are not in
${\cal A}_{\rm univ}^{C^*}$, but they are formal
(unbounded) limits of elements in
${\cal A}_{\rm univ}^{C^*}$.
Mack and Schomerus \cite{MS1,MS2} presented
endomorphisms $\varrho_J$ such that
$\pi_0\circ\varrho_J\cong\pi_J$, $J=\frac{1}{2},1$ is
fulfilled. But, as already mentioned, these endomorphisms are
not localized, i.e. there is no interval $I'\neq\emptyset$
such that their action is trivial on ${\cal A}(I')$. In this
paper we present examples $\varrho_{1/2}^{\rm loc}$,
$\varrho_1^{\rm loc}$ of localized endomorphisms which are
unitarily equivalent to those global ones in composition
with the vacuum representation. However, our construction
slightly differs from the formalism used by
Mack and Schomerus.

Since the set ${\cal J}$ of proper subintervals
on the circle is not directed, a global algebra
cannot be defined as the $C^*$-inductive
limit of the system $\{{\cal A}(I),I\in{\cal J}\}$.
The global algebra ${\cal A}_{\rm univ}^{C^*}$
has to be considered as the algebra generated
{\sl freely} by all local algebras
${\cal A}(I)$. This is the universal algebra
(but generated by local $C^*$-algebras instead
of von Neumann algebras) in the sense of
Fredenhagen, Rehren and Schroer \cite{FRS2,Fred3}.
Its center is nontrivial, generated by a unitary
element $Y$. The  $C^*$-algebra of the punctured
circle ${\cal A}(I_\zeta)$ where $I_\zeta=S^1
\setminus\{\zeta\}$, $\zeta\in S^1$ an arbitrary point,
has a trivial center and ${\cal A}_{\rm univ}^{C^*}$
is generated by ${\cal A}(I_\zeta)$ and $Y$.
The local algebras ${\cal A}(I)$ are even subalgebras
of (selfdual) CAR algebras over spaces $L^2(I)$.
Also the global observable algebra
${\cal A}_{\rm univ}^{C^*}$ is the even subalgebra
of a global field algebra, the universal
Majorana algebra {\sf Maj}. It has the structure of
the direct sum of two selfdual
CAR algebras over $L^2(S^1)$. Alternatively, it may be
regarded as the algebra which is generated by an
anticommuting universal Majorana field living on the
double cover $\tilde{S}^1$ of the unit circle
\cite{MS1,MS2}. For recovering the local algebras
${\cal A}(I)$ as even subalgebras of {\sf Maj} by
explicit construction, we have to fix an
arbitrary reference point $\zeta$ (``point at
infinity'') on the circle.

The non-trivial center of the global algebra
${\cal A}_{\rm univ}^{C^*}$ implies that its irreducible
representations can no longer be faithful. This
leads to some deviations from the customary DHR
program. In particular, the vacuum representation $\pi_0$
of ${\cal A}_{\rm univ}^{C^*}$ cannot be faithful.
There is another difference between the formalism used
by Mack and Schomerus and the common DHR framework:
The local algebras ${\cal A}(I)$ are defined as
$C^*$-algebras instead of von Neumann algebras.
But the use of von Neumann algebras is crucial
for the analysis of statistics and fusion. On the
other hand, if one works with local von Neumann
algebras  (weak closures of
${\cal A}(I)$ in the vacuum representation), the
universal algebra becomes even larger, in particular,
its center is larger than that of
${\cal A}_{\rm univ}^{C^*}$,
generated by $Y$. Such a universal algebra appears
to be hard to handle.

During our investigations, it
turned out to be much more comfortable to formulate the
theory on the punctured circle. Such a formulation is
possible because in our theory Haag duality remains
valid on the punctured circle. Having fixed a ``point
at infinity'' $\zeta\in S^1$, the set
${\cal J}_\zeta$ of those open intervals
such that $\zeta$ is not contained in their
closures (``finite intervals'') is directed
by inclusion. So the theory can be developed
close to the DHR program. We define local von
Neumann algebras ${\cal R}(I)=\pi_0({\cal A}(I))''$,
and the net $\{{\cal R}(I),I\in{\cal J}_\zeta\}$
generates a global $C^*$-algebra $\mathfrak{A}_\zeta$
in the natural way; $\mathfrak{A}_\zeta$ may be
regarded as the algebra of quasilocal observables.
The representations $\pi_J$ and the
localized endomorphisms $\varrho_J^{\rm loc}$
we present are at first defined on local
$C^*$-algebras ${\cal A}(I)$. We show that they
possess an extension to the net of von Neumann
algebras. A local normality relation is used
for the proof. Using some results of CAR theory,
we establish that indeed unitary equivalence
$\pi_0\circ\varrho_J^{\rm loc}\cong\pi_J$ holds.
Finally, we manifestly prove the Ising fusion
rules in terms of equivalence classes of localized
endomorphisms, and we construct statistics
operators and a left inverse.

With respect to the proof of fusion rules, we
believe to close a gap left in \cite{MS1,MS2}.
Mack and Schomerus had already proven that their
global endomorphisms obey Ising fusion rules.
But caused by the use of local $C^*$-algebras and
a non-faithful vacuum representation, this result
could not be generalized to all endomorphisms,
which lead to equivalent representations. (There
are counterexamples.) The existence of unitary
intertwiners in the observable algebra, being
essential for such a generalization, is not
guaranteed if one does not work with von Neumann
algebras. However, establishing a theory based
on local von Neumann algebras on the punctured
circle, we close the gap. But it should be
mentioned that there exist also successful
methods for proving conformal field theory
fusion rules without the use of localized
endomorphisms, e.g.~\cite{Loke,Wass}.

Our paper is organized as follows. We present
Majorana fields, local $C^*$-algebras of observables
and the global algebra ${\cal A}_{\rm univ}^{C^*}$ in
Section 2. Using some ideas of Szlach\'{a}nyi
\cite{Szl2}, we discuss the origin of its central
element $Y$. We introduce the universal Majorana
algebra ${\sf Maj}$,  we describe the representation
theory of ${\sf Maj}$ and ${\cal A}_{\rm univ}^{C^*}$
and we introduce the Mack-Schomerus endomorphisms.
Section 3 begins with a brief recapitulation of
the CAR theory and some results we need. Next we
describe the representation theory of ${\sf Maj}$
and ${\cal A}_{\rm univ}^{C^*}$ in view of Araki's
selfdual CAR algebra and quasifree states. We
discuss the restriction to the algebra
${\cal A}(I_\zeta)$ of the punctured circle.
Then we introduce our examples of localized
endomorphisms and we analyze the induced
representations. In Section 4 we discuss the
extension of representations and localized
endomorphisms to local von Neumann algebras and
to the global algebra $\mathfrak{A}_\zeta$. In this
framework, we prove fusion rules and give
statistics operators and a left inverse.

\section{Algebras, Representations and global
Endomorphisms of the Chiral Ising Model}
In this section, we develop and analyze the
formalism used by Mack and Schomerus to
describe the chiral Ising model.
\subsection{Local $C^*$-Algebras and their
Universal Algebra}
We begin our investigations with a brief description
of the field algebra, the local and the global
observable algebras of the chiral Ising model.
Our starting point is a Majorana field $\psi$
on the unit circle $S^1$ which has anticommutation
relations
\begin{equation}
\{ \psi(z)^*,\psi(w) \} = 2\pi\I z \delta(z-w)
\end{equation}
and hermiticity condition
\begin{equation}
\psi(z)^* = z \psi(z).
\end{equation}
We consider smeared fields
\begin{equation}
\psi(f) = \oint_{S^1} \frac{dz}{2\pi\I z^\frac{1}{2}}
f(z) \psi(z), \qquad f(z) \in L^2(S^1).
\end{equation}
These objects obey the canonical anticommutation relations
(CAR) of the ca\-nonical generators of Araki's \cite{Ara1,Ara2}
selfdual CAR-algebra ${\cal C}({\cal K},\Gamma)$ over the
Hilbert space ${\cal K}=L^2(S^1)$ with the antiunitary
involution $\Gamma$ of complex conjugation. We have
\begin{equation}
\{ \psi(f)^*,\psi(g) \} = \langle f,g \rangle {\bf 1}
\end{equation}
with
\begin{equation}
\psi(f)^* = \psi(\Gamma f), \qquad \langle f,g \rangle =
\oint_{S^1} \frac{dz}{2\pi\I z} \overline{f(z)} g(z).
\end{equation}
As local algebras ${\cal A}(I)$ with some open interval
$I\subset S^1$, $I\neq S^1$ we define those unital
algebras which are generated by bilinear expressions
\[ B_I(f,g) = \psi(f)\psi(g), \qquad \mbox{supp}(f)
\subset I, \mbox{supp}(g) \subset I \]
in the Majorana fields. These generators are complex
linear in both arguments and obey relations
\begin{eqnarray}
\label{Bbegin}
2 B_I (f,f) &=& \langle \Gamma f,f \rangle {\bf 1},\\
2 B_I (f,g) B_I (g,h) &=& \langle \Gamma g,g \rangle
B_I (f,h), \\
B_I (f,g)^* &=& B_I (\Gamma g,\Gamma f),
\end{eqnarray}
where $f,g,h\in L^2(S^1)$ are functions with support
in $I$. Next we consider the algebras ${\cal A}(I)$
as defined only by these abstract relations.
Since the set ${\cal J}$ of open, non-void intervals
$I\neq S^1$ on the circle is not directed there is no
inductive limit for the algebras ${\cal A}(I),
I\in{\cal J}$. But with the additional relation
\begin{equation}
\label{Bend}
B_I(f,g) = B_J(f,g), \qquad I \subset J
\end{equation}
one can construct a global algebra
${\cal A}_{\rm univ}^{C^*}$ which is
generated by all $B_I(f,g)$, $f,g\in L^2(S^1)$ and
$I\in {\cal J}$ \cite{Szl2,Fred3,ich}.
Perhaps one could expect that the result is the
even subalgebra of the selfdual CAR algebra over
the whole circle $S^1$. We will show that this is
actually not the case; instead there occurs a
central element $Y\in{\cal A}_{\rm univ}^{C^*}$ which
will finally lead to the fact that
${\cal A}_{\rm univ}^{C^*}$ is the direct sum of
two of those even CAR algebras. Let now $I_1$ and $I_2$ be
two disjoint intervals and let $J_+$ and $J_-$ be intervals
containing both of them, one from the left side and one
from the right side, so that $J_+ \cup J_-=S^1$.
Choose real functions $f_j\in L^2(S^1)$ with
$\|f_j\|^2=2$ with supp$(f_j)\subset I_j$, $j=1,2$.
Then define
\begin{equation}
\label{Y}
Y = B_{J_+}(f_1,f_2) B_{J_-}(f_2,f_1).
\end{equation}
One finds that $Y$ is unitary, self-adjoint and independent
of the special choice of $f_1,f_2,I_1,I_2,J_+,J_-$.
Moreover, $Y$ is in the center of ${\cal A}_{\rm univ}^{C^*}$.
For every $\zeta\in S^1$ and $I_\zeta=S^1\setminus
\{\zeta\}$, the global algebra ${\cal A}_{\rm univ}^{C^*}$ is
generated by ${\cal A}(I_\zeta)$ and $Y$
\cite{Szl2,Fred3,ich}.

We now want to reconstruct the global, or, ``universal''
algebra ${\cal A}_{\rm univ}^{C^*}$ by a global field
algebra, the universal Majorana algebra.
\begin{definition}
\label{Maj}
The universal Majorana algebra {\sf Maj} is defined as the
direct sum of the selfdual CAR algebra over
$(L^2(S^1),\Gamma)$ with itself, i.e.
\begin{equation}
\mbox{\sf Maj}= {\cal C}(L^2(S^1),\Gamma) \oplus
{\cal C}(L^2(S^1),\Gamma).
\end{equation}
The center of {\sf Maj} is generated by the element
\begin{equation}
Y= (-{\bf 1}) \oplus {\bf 1}
\end{equation}
and we have the two subalgebras
\begin{equation}
\mbox{\sf Maj}_{\rm NS}=\frac{1}{2}({\bf 1}-Y)\mbox{\sf Maj},
\qquad \mbox{\sf Maj}_{\rm R}=\frac{1}{2}({\bf 1}+Y)\mbox{\sf Maj}.
\end{equation}
\end{definition}
The universal Majorana algebra is a well defined
$C^*$-algebra since ${\cal C}(L^2(S^1),\Gamma)$ is. For
clarifying the connection between our definition and
the definition of {\sf Maj} given by Mack and Schomerus
\cite{MS1} we consider the following two orthonormal
bases of $L^2(S^1)$
\[ \left\{e_r, r\in\mathbb{Z}+\frac{1}{2}\right\} \qquad
\mbox{and} \qquad \{e_n, n\in\mathbb{Z} \}, \]
where $e_a(z)=z^a$ for $z=\E^{\I\phi}\in S^1$,
$-\pi<\phi\le\pi$, $a\in\frac{1}{2}\mathbb{Z}$.
We define the elements of {\sf Maj} (Fourier modes)
\begin{eqnarray*}
b_r &=& \psi(e_r) \oplus 0 , \qquad r\in\mathbb{Z}+\frac{1}{2},\\
b_n &=& 0 \oplus \psi(e_n) , \qquad n\in\mathbb{Z}.
\end{eqnarray*}
Then we have
\begin{itemize}
\item $\mbox{\sf Maj}_{\rm NS}$ is generated by the modes
$b_r$, $r\in \mathbb{Z}+\frac{1}{2}$,
\item $\mbox{\sf Maj}_{\rm R}$ is generated by the modes
$b_n$, $n\in \mathbb{Z}$,
\item {\sf Maj} is generated by the modes
$b_a$, $a\in \frac{1}{2}\mathbb{Z}$,
\end{itemize}
and the Fourier modes satisfy relations
\begin{equation}
\{ b_a,b_c \} = \frac{1}{2} ({\bf 1}+(-1)^{2a}Y) \delta_{a,-c},
\quad b_a^*=b_{-a},
\end{equation}
\begin{equation}
Yb_a=(-1)^{2a}b_a,\quad [Y,b_a]=0,\quad Y=Y^*,\quad Y^2={\bf 1}.
\end{equation}
It is convenient to understand the elements of {\sf Maj} as smeared
fields as well. We define the Hilbert space
\[ \hat{\cal K} = L^2(S^1) \oplus L^2(S^1) \]
which may be identified with $L^2({\tilde{S}}_1)$, where
${\tilde{S}}_1$ denotes the double cover of $S^1$.
Hence each element $\hat{f}\in\hat{\cal K}$ has the
unique decomposition
\[ \hat{f} = f_{\rm NS} \oplus f_{\rm R}, \qquad f_{\rm NS},f_{\rm R}
\in L^2(S^1). \]
On $\hat{\cal K}$ we have the antiunitary involution
\[ \hat{\Gamma}=\Gamma \oplus \Gamma. \]
We define the field $\hat{\psi}(\hat{f})\in${\sf Maj} by
\[ \hat{\psi}(\hat{f}) = \psi(f_{\rm NS}) \oplus \psi(f_{\rm R}),\]
so that we have the conjugation
\[ \hat{\psi}(\hat{f})^* =
\hat{\psi}(\hat{\Gamma}\hat{f}),\]
anticommutation relations
\[ \{ \hat{\psi}(\hat{f})^*,\hat{\psi}(\hat{g}) \} =
\frac{1}{2} ({\bf 1}-Y) \langle f_{\rm NS}, g_{\rm NS} \rangle
+ \frac{1}{2} ({\bf 1}+Y) \langle f_{\rm R}, g_{\rm R}
\rangle, \]
boundary condition
\[ Y \hat{\psi}(\hat{f}) = \hat{\psi}(y\hat{f}),
\qquad y=(-{\bf 1})\oplus {\bf 1} \in
{\cal B}(\hat{\cal K}) \]
and
\[ [ Y , \hat{\psi}(\hat{f}) ] = 0. \]
We now want to redefine the local generators $B_I(f,g)
\in{\cal A}(I)$ as even elements of {\sf Maj}. For that we
have to fix an arbitrary point $\zeta\in S^1$. We
distinguish between two cases:

\noindent {\sl Case 1}: For all intervals $I\in{\cal J}$
with $\zeta\notin I$ we set
\begin{equation}
\label{C1}
B_I(f,g) = \hat{\psi}(\hat{f})\hat{\psi}(\hat{g}),
\qquad \hat{f}=f\oplus f \in \hat{\cal K},\quad
\hat{g}=g\oplus g \in \hat{\cal K}.
\end{equation}

\noindent {\sl Case 2}: For every interval $I\in{\cal J}$
with $\zeta\in I$ the point $\zeta$ splits $I$ in two
disjoint intervals $I_1$ and $I_2$ so that $I=I_1 \cup
\{\zeta\} \cup I_2$. Let $\chi_j$ be the characteristic
functions of $I_j$ and set $f_j=\chi_j f$, $g_j=\chi_j g$,
$j=1,2$. Then we set
\begin{eqnarray}
\label{C2}
B_I(f,g) &=& \hat{\psi} ({\hat{f}}_1) \hat{\psi}
({\hat{g}}_1) + \hat{\psi} ({\hat{f}}_2) \hat{\psi}
({\hat{g}}_2) + Y \hat{\psi} ({\hat{f}}_1) \hat{\psi}
({\hat{g}}_2) + Y \hat{\psi} ({\hat{f}}_2) \hat{\psi}
({\hat{g}}_1), \\
&&{\hat{f}}_j = f_j \oplus f_j \in \hat{\cal K},\quad
{\hat{g}}_j =g_j \oplus g_j \in \hat{\cal K}, \qquad
j=1,2. \nonumber
\end{eqnarray}
It is an easy but less beautiful work to control that
these $B_I(f,g)$ satisfy the relations (\ref{Bbegin}) -
(\ref{Bend}), and
that relation (\ref{Y}) is fulfilled with the $Y$ of
Definition \ref{Maj}, also independent of the functions and
intervals \cite{ich}. It is not hard to see that the
identifications (\ref{C1}),(\ref{C2}) define an
isomorphism between ${\cal A}_{\rm univ}^{C^*}$ and
the even part ${\sf Maj}^+$ of ${\sf Maj}$, too \cite{Szl2}.
Thus we are allowed to identify the global observable algebra
${\cal A}_{\rm univ}^{C^*}$ with the even part of {\sf Maj}.

\subsection{Representations and Endomorphisms}
Each of the algebras $\mbox{\sf Maj}_{\rm NS}$ and
$\mbox{\sf Maj}_{\rm R}$ possesses a faithful cyclic
representation
$({\cal H}_{\rm NS},\pi_{\rm NS},|\Omega_{\rm NS}\rangle)$ and
$({\cal H}_{\rm R},\pi_{\rm R},|\Omega_{\rm R}\rangle)$ which is
characterized by
\begin{eqnarray}
\label{defNS}
\pi_{\rm NS} (b_r)|\Omega_{\rm NS}\rangle &=& 0, \quad r>0,
\quad r\in \mathbb{Z}+\frac{1}{2}, \\
\label{defR}
\pi_{\rm R} (b_n) |\Omega_{\rm R} \rangle&=& 0, \quad n>0,
\quad n\in \mathbb{Z},
\end{eqnarray}
respectively.
The NS-representation is uniquely characterized
(all matrix-ele\-ments can be computed and the vector
$|\Omega_{\rm NS}\rangle$ is defined to be cyclic).
In the R-representation, the action of the
self-adjoint $b_0$ on the cyclic $|\Omega_{\rm R}\rangle$
is not completely fixed. To determine the
R-representation uniquely too, we require in
addition that the vectors $|\Omega_{\rm R}\rangle$ and
$\pi_{\rm R}(b_0)|\Omega_{\rm R}\rangle$ are orthogonal in
${\cal H}_{\rm R}$,
\begin{equation}
\label{ortho}
\langle \Omega_{\rm R} | \pi_{\rm R} (b_0)
| \Omega_{\rm R} \rangle = 0.
\end{equation}
One can consider these representations as those of {\sf Maj}
on the space ${\cal H}_{\rm NS}\oplus{\cal H}_{\rm R}$ by the
requirement
\begin{equation}
\pi_{\rm NS}(Y)= -{\bf 1}, \qquad \pi_{\rm R} (Y) = {\bf 1}
\end{equation}
which leads automatically to
\begin{equation}
\label{def0}
\pi_{\rm NS}(b_n)=0,\qquad n\in \mathbb{Z} \qquad \mbox{and}
\qquad \pi_{\rm R}(b_r)=0,\qquad r\in \mathbb{Z}+\frac{1}{2}
\end{equation}
i.e. $\pi_{\rm NS}$ lives only on $\mbox{\sf Maj}_{\rm NS}$ and
$\pi_{\rm R}$ on $\mbox{\sf Maj}_{\rm R}$. Of course, both
representations are then no longer faithful.
The NS-representation is irreducible, the
R-representation is not; it decomposes into two
irreducible subrepresentations
$({\cal H}_{\rm R}^+,\pi_{\rm R}^+)$ and $({\cal H}_{\rm R}^-,\pi_{\rm R}^-)$
(see below) which are generated by the
action of $\pi_{\rm R}$({\sf Maj})
on vectors $|\Omega_{\rm R}^+\rangle$ and
$|\Omega_{\rm R}^-\rangle$, respectively, where
\[ |\Omega_{\rm R}^\pm\rangle = \frac{1}{\sqrt{2}}
|\Omega_{\rm R}\rangle \pm \pi_{\rm R}(b_0)|\Omega_{\rm R}\rangle. \]
These states are eigenstates of $\pi_{\rm R}(b_0)$
with eigenvalues $\pm 2^{-\frac{1}{2}}$
\cite{Szl1}. We are now interested in what
happens, when the representations of {\sf Maj},
$\pi_{\rm NS}$ and $\pi_{\rm R}$, are restricted to the
observable algebra which is the even subalgebra
of {\sf Maj},
${\cal A}_{\rm univ}^{C^*}=\mbox{\sf Maj}^+$.
It is known that the NS-representation splits into
two irreducibles,
\begin{equation}
\pi_{\rm NS}|_{{\cal A}_{\rm univ}^{C^*}} =
\pi_0 \oplus \pi_1,\qquad
{\cal H}_{\rm NS}={\cal H}_0 \oplus {\cal H}_1,
\end{equation}
and the R-representation decomposes into two
equivalent ones,
\begin{equation}
\pi_{\rm R}|_{{\cal A}_{\rm univ}^{C^*}} =
\pi_{1/2} \oplus \pi_{1/2}',\qquad {\cal H}_{\rm R} =
{\cal H}_{1/2} \oplus {\cal H}_{1/2}'.
\end{equation}
The subspaces ${\cal H}_0$, ${\cal H}_1$,
${\cal H}_{1/2}$ and ${\cal H}_{1/2}'$
are spanned by vectors
\begin{eqnarray*}
\pi_{\rm NS}(b_{-r_{2N}} \cdots b_{-r_1}) |\Omega_{\rm NS}\rangle &\in&
{\cal H}_0,\,\, r_i\in \mathbb{N}_0+\frac{1}{2},\,\, r_{2N}>
\cdots >r_1,\\
\pi_{\rm NS}(b_{-r_{2N+1}} \cdots b_{-r_1}) |\Omega_{\rm NS}\rangle &\in&
{\cal H}_1,\,\,r_i\in \mathbb{N}_0+\frac{1}{2},\,\, r_{2N+1}>
\cdots >r_1,\\
\pi_{\rm R}(b_{-n_{2N}} \cdots b_{-n_1}) |\Omega_{\rm R}\rangle &\in&
{\cal H}_{1/2}, \quad n_i\in \mathbb{N}_0,\,\, n_{2N}>
\cdots >n_1,\\
\pi_{\rm R}(b_{-n_{2N+1}} \cdots b_{-n_1}) |\Omega_{\rm R}\rangle &\in&
{\cal H}_{1/2}', \quad n_i\in \mathbb{N}_0,\,\, n_{2N+1}>
\cdots >n_1,
\end{eqnarray*}
with $N\in \mathbb{N}_0$. We remark that the subspaces
${\cal H}_{1/2}$ and ${\cal H}_{1/2}'$
do not coincide with ${\cal H}_{\rm R}^+$ and ${\cal H}_{\rm R}^-$.
How is that possible? The reason is that the
subrepresentations $\pi_{\rm R}^+$ and $\pi_{\rm R}^-$, when
restricted to the observable algebra
${\cal A}_{\rm univ}^{C^*}$, become equivalent
\cite{Szl1}, and see below.  Therefore the
decomposition into invariant subspaces is not
unique.

Mack and Schomerus \cite{MS1,MS2} defined the following
endomorphisms of {\sf Maj} which restrict to endomorphisms
of the global observable algebra ${\cal A}_{\rm univ}^{C^*}$.
\begin{definition}
\label{endo}
The endomorphisms $\varrho_J$,
$J=0,\frac{1}{2},1$ of {\sf Maj} are defined by their
action on the generators as follows,
\begin{eqnarray}
\varrho_0 = id ,\qquad \qquad \qquad \qquad \qquad \qquad\\
\varrho_{1/2}(b_a) = \left\{ \begin{array}{cl}
\I b_{a+\frac{1}{2}} & \quad a \ge \frac{1}{2} \\
\frac{\I}{\sqrt{2}} (b_\frac{1}{2}-b_{-\frac{1}{2}}) & \quad a=0 \\
- \I b_{a-\frac{1}{2}} & \quad a\le -\frac{1}{2} \end{array}\right.
,\qquad \varrho_{1/2}(Y)=-Y ,\\
\varrho_1(b_a)= \left\{ \begin{array}{cl} -b_a & \quad a \neq
0, \pm \frac{1}{2} \\
b_{-a} & \quad a = 0, \pm \frac{1}{2}  \end{array} \right.
,\qquad \varrho_1(Y)=Y .\quad
\end{eqnarray}
\end{definition}
It is shown \cite{MS1} that these endomorphisms fulfill
\begin{eqnarray}
\pi_{\rm NS} \circ \varrho_{1/2} &\cong& \pi_{\rm R},\\
\label{rho1}
\pi_{\rm NS} \circ \varrho_1 &\cong& \pi_{\rm NS},\\
\pi_0 \circ \varrho_J &\cong& \pi_J, \qquad J=0,
\frac{1}{2},1,
\end{eqnarray}
where relation (\ref{rho1}) is the most trivial one because
$\varrho_1$ is inner in {\sf Maj}, implemented by the
unitary self-adjoint
\[ R=\sqrt{2}b_0+b_\frac{1}{2}+b_{-\frac{1}{2}}
\in \mbox{\sf Maj}. \]
We can define these endomorphisms by the formula
\[ \varrho_J(\hat{\psi}(\hat{f}))=
\hat{\psi}(\hat{V}_J\hat{f}) \]
where $\hat{V}_J$ are the following isometries on
$\hat{\cal K}=L^2(S^1)\oplus L^2(S^1)$,
\begin{equation}
\label{matrix}
{\hat{V}}_0=\left( \begin{array}{cc}
{\bf 1} & 0 \\ 0 & {\bf 1}
\end{array} \right),\quad
{\hat{V}}_{1/2} =\left( \begin{array}{cc} 0 &
V_{1/2} \\ V_{1/2}' & 0
\end{array} \right),\quad
{\hat{V}}_1=\left( \begin{array}{cc} V_1 & 0 \\ 0 & V_1'
\end{array} \right);
\end{equation}
the isometries (Bogoliubov operators, see below)
$V_{1/2},V_{1/2}',V_1,V_1'\in{\cal B}(L^2(S^1))$
are defined by
\begin{eqnarray*}
V_{1/2} &=& \frac{\I}{\sqrt{2}}\Big(|e_\frac{1}{2}
\rangle\langle e_0|-|e_{-\frac{1}{2}}\rangle
\langle e_0|\Big) + \I \sum_{n=1}^\infty \Big(
|e_{n+\frac{1}{2}}\rangle\langle
e_n|-|e_{-n-\frac{1}{2}}\rangle\langle
e_{-n}|\Big),  \\
V_{1/2}' &=&  \I \sum_{n=1}^\infty \Big( |e_n
\rangle\langle e_{n-\frac{1}{2}}|-|e_{-n} \rangle
\langle e_{-n+\frac{1}{2}}| \Big),\\
V_1 &=&  |e_\frac{1}{2}\rangle\langle e_{-\frac{1}{2}}|+
|e_{-\frac{1}{2}}\rangle\langle e_\frac{1}{2}|
- \sum_{n=1}^\infty \Big(|e_{n+\frac{1}{2}}\rangle\langle
e_{n+\frac{1}{2}}|+|e_{-n-\frac{1}{2}}\rangle\langle
e_{-n-\frac{1}{2}}| \Big), \\
V_1' &=& |e_0\rangle\langle e_0| -\sum_{n=1}^\infty
\Big( |e_n\rangle\langle e_n|+|e_{-n}\rangle\langle
e_{-n}| \Big).
\end{eqnarray*}
It is worthy of note that the two non-vanishing entries,
each in $\hat{V}_{1/2}$ and $\hat{V}_1$, are
actually different.

\section{Localized Endomorphisms}
In this section we present our localized
endomorphisms in terms of Bogoliubov transformations.
After a brief summary of mathematical results which
we will use, we introduce them as endomorphisms of the
algebra of the punctured circle.

\subsection{The Selfdual CAR Algebra: Some
Useful Results}
For a better handling of our techniques we give a brief
repetition of Araki's selfdual CAR algebra
${\cal C}({\cal K},\Gamma)$ and quasifree
states \cite{Ara1,Ara2}.
We consider a Hilbert space ${\cal K}$ with an
antiunitary involution $\Gamma$ (complex conjugation),
$\Gamma^2={\bf 1}$, which fulfills
\[ \langle \Gamma f, \Gamma g \rangle = \langle g ,
f \rangle, \qquad f,g\in{\cal K}. \]
The selfdual CAR algebra ${\cal C}({\cal K},\Gamma)$
is defined to be the $C^*$-norm closure of the algebra
which is generated by the image
of a linear mapping $\psi$ which maps elements
$f\in{\cal K}$ to canonical generators $\psi(f)$,
so that
\[ \psi(f)^*=\psi(\Gamma f),\qquad
\{ \psi(f)^*,\psi(g) \} = \langle f,g \rangle {\bf 1} \]
holds. The $C^*$-norm satisfies \cite{Ara2}
\[ \|\psi(f)\| = \frac{1}{\sqrt{2}} \sqrt{\|f\|^2
+ \sqrt{\|f\|^4-|\langle f,\Gamma f \rangle |^2 }}. \]
In particular, we have the inequality
\begin{equation}
\label{Cstern}
\|\psi(f)\| \le \|f\|.
\end{equation}
Elements of the set
\[ {\cal I}({\cal K},\Gamma) = \{ V\in {\cal B}({\cal K})
\,|\, [V,\Gamma]=0,\,\, V^*V={\bf 1} \} \]
of $\Gamma$ commuting isometries on ${\cal K}$ are
called Bogoliubov operators. Every Bogoliubov
operator $V\in {\cal I}({\cal K},\Gamma)$ defines an
endomorphism $\varrho_V$ of ${\cal C}({\cal K},\Gamma)$,
defined by its action on the canonical generators,
\[ \varrho_V(\psi(f))=\psi(Vf). \]
Moreover, if $V\in {\cal I}({\cal K},\Gamma)$ is surjective
(i.e. unitary), then $\varrho_V$ is an automorphism.
\begin{definition}
A state $\omega$ of ${\cal C}({\cal K},\Gamma)$ is called
quasifree if for all $n\in \mathbb{N}$
\begin{eqnarray}
\omega (\psi(f_1) \cdots \psi(f_{2n+1})) &=& 0, \\
\label{perm1}
\omega (\psi(f_1) \cdots \psi(f_{2n})) &=&
(-1)^\frac{n(n-1)}{2} \sum_\sigma
\mbox{\rm sign} \sigma\prod_{j=1}^n \omega
(\psi(f_{\sigma(j)})\psi(f_{\sigma(n+j)}))\,\,\,
\end{eqnarray}
holds. The sum runs over all permutations
$\sigma \in {\cal S}_{2n}$ with the property
\begin{equation}
\label{perm2}
\sigma (1) < \sigma (2) < \cdots < \sigma (n),
\qquad \sigma (j) < \sigma (j+n), \qquad j=1,\ldots,n.
\end{equation}
\end{definition}
Quasifree states are therefore completely characterized by
their two point function. It is known that there is a
one to one correspondence between the set of quasifree
states and the set
\[ {\cal Q}({\cal K},\Gamma)=\{S\in{\cal B}({\cal K})
\,|\, S=S^*,\, 0 \le S \le {\bf 1},\,
S + \Gamma S \Gamma = {\bf 1} \}, \]
given by the formula
\begin{equation}
\label{phi-S}
\omega (\psi(f)^* \psi(g)) = \langle f,Sg \rangle .
\end{equation}
The quasifree state characterized by Eq.~(\ref{phi-S})
is denoted
by $\omega_S$. A quasifree state, composed with a
Bogoliubov endomorphism is again a quasifree state,
namely we have $\omega_S\circ\varrho_V=
\omega_{V^*SV}$. The projections in ${\cal Q}({\cal K},\Gamma)$
are called basis projections. If $P$ is a basis projection
then the state $\omega_P$ is pure and is called a
Fock state. The corresponding GNS representation
$({\cal H}_P,\pi_P,|\Omega_P\rangle)$
is irreducible, it is called the Fock representation;
the vector $|\Omega_P\rangle\in{\cal H}_P$ is
called the Fock vacuum. Araki proved
\cite{Ara1,Ara2} that a state $\omega$
of ${\cal C}({\cal K},\Gamma)$ which satisfies
\begin{equation}
\omega(\psi(f)\psi(f)^*)=0, \qquad f\in P{\cal K}
\end{equation}
for a basis projection $P$ is automatically
the Fock state $\omega=\omega_P$.

We now come to an important quasiequivalence
criterion for quasifree states. It was developed for
the case of gauge invariant quasifree states by Powers
and St{\o}rmer \cite{Pow} and generalized for arbitrary
quasifree states by Araki \cite{Ara1}. Unitary equivalence
(denoted by "$\cong$") or quasiequivalence
(denoted by "$\approx$") of states means always that
the corresponding GNS representations are unitarily
equivalent or quasiequivalent, respectively.
\begin{theorem}
\label{Araki}
Two quasifree states $\omega_{S_1}$ and
$\omega_{S_2}$ of ${\cal C}({\cal K},\Gamma)$
are quasiequivalent if and only if
\begin{equation}
S_1^\frac{1}{2} - S_2^\frac{1}{2} \in {\cal J}_2 ({\cal K}),
\end{equation}
where ${\cal J}_2({\cal K})$ denotes the ideal of
Hilbert Schmidt operators in ${\cal B}({\cal K})$.
\end{theorem}
We now can conclude that two Fock states
$\omega_{P_1}$ and $\omega_{P_2}$ are unitarily equivalent,
if and only if $P_1-P_2$ is Hilbert Schmidt class, or,
if $\omega_P$ is a Fock state and $\varrho_V$
is a Bogoliubov endomorphism, that
$\omega_P\circ\varrho_V\approx\omega_P$ if and only if
$P-(V^*PV)^\frac{1}{2}$
is Hilbert Schmidt class. But in most cases we study
{\sl representations} of the form $\pi_P\circ\varrho_V$,
where $\varrho_V$ is a Bogoliubov endomorphism and
$\pi_P$ a Fock representation of ${\cal C}({\cal K},\Gamma)$.
Such a representation $\pi_P\circ\varrho_V$ is in general
not cyclic but it is equivalent to a multiple of the
GNS representation $\pi_{V^*PV}$ of the state
$\omega_{V^*PV}=\omega_P\circ\varrho_V$. The
multiplicity is given by $2^{N_V}$ where $N_V$
is the dimension of the intersection of
ker$V^*$ and $P{\cal K}$ \cite{Binnenneu,Rideau},
i.e.
\begin{equation}
\label{cyclics}
\pi_P\circ\varrho_V\cong 2^{N_V}\pi_{V^*PV},
\qquad N_V={\rm dim}({\rm ker}V^*\cap P{\cal K}).
\end{equation}
This is a decomposition of $\pi_P\circ\varrho_V$
into cyclic subrepresentations but in general
not into irreducibles. A decomposition into
irreducibles is provided by the following theorem
which was proven in \cite{Decom} and, in a
different way, in \cite{Binnenneu}.
\begin{theorem}
\label{evenodd}
Let $V$ be a Bogoliubov operator with
$M_V=\mbox{\rm dim ker}V^* < \infty$. If $M_V$ is
an even integer we have (with notations as above)
\begin{equation}
\pi_P\circ\varrho_V \cong 2^\frac{M_V}{2} \pi_{P'}
\end{equation}
where $\pi_{P'}$ is an (irreducible) Fock representation.
If $M_V$ is odd then we have
\begin{equation}
\label{decomodd}
\pi_P\circ\varrho_V \cong 2^\frac{M_V-1}{2}
(\pi_+ \oplus \pi_-)
\end{equation}
where $\pi_+$ and $\pi_-$ are mutually inequivalent,
irreducible representations.
\end{theorem}
The representations $\pi_\pm$ occurring in Eq.
(\ref{decomodd}) are called pseudo Fock
representations \cite{Ara1}.
Consider the automorphism $\alpha_{-1}$ of
${\cal C}({\cal K},\Gamma)$ which is defined by
$\alpha_{-1}(\psi(f))=-\psi(f)$. We define the
even algebra ${\cal C}({\cal K},\Gamma)^+$
to be the subalgebra of $\alpha_{-1}$-fixpoints,
\begin{equation}
{\cal C}({\cal K},\Gamma)^+ = \{ x \in
{\cal C}({\cal K},\Gamma) | \alpha_{-1}(x)=x \}.
\end{equation}
We now are interested in what happens when our
representations of ${\cal C}({\cal K},\Gamma)$
are restricted to the even algebra. For basis
projections $P_1,P_2$, with $P_1-P_2$ Hilbert Schmidt
class, Araki and D.E.~Evans \cite{AE} defined
an index, taking values $\pm 1$,
\[ \mbox{ind}(P_1,P_2)= (-1)^{{\rm dim}(
P_1{\cal K}\cap({\bf 1}-P_2){\cal K})}. \]
The automorphism $\alpha_{-1}$ leaves any
quasifree state $\omega_S$ invariant. Thus
$\alpha_{-1}$ is implemented in $\pi_S$. In particular,
in a Fock representation $\pi_P$, $\alpha_{-1}$
extends to an automorphism $\bar{\alpha}_{-1}$ of
$\pi_P({\cal C}({\cal K},\Gamma))''=
{\cal B}({\cal H}_P)$. The following proposition
is taken from \cite{Ara2}.
\begin{proposition}
\label{ind}
Let $U\in{\cal I}({\cal K},\Gamma)$ be a unitary
Bogoliubov operator and let $P$ be a basis
projection such that $P-U^*PU$ is Hilbert-Schmidt
class. Denote by $Q(U)\in{\cal B}({\cal H}_P)$ the
unitary which implements $\varrho_U$ in $\pi_P$. Then
\begin{equation}
\bar{\alpha}_{-1}(Q(U)) = \sigma(U) Q(U),\qquad
\sigma(U)=\pm 1.
\end{equation}
In particular,
\begin{equation}
\sigma(U) = \mbox{\rm ind}(P,U^*PU).
\end{equation}
\end{proposition}
Furthermore, one has \cite{AE,Ara2}
\begin{theorem}
\label{resteven}
Restricted to the even algebra
${\cal C}({\cal K},\Gamma)^+$, a Fock representation
$\pi_P$ splits into two mutually inequivalent, irreducible
subrepresentations,
\begin{equation}
\pi_P|_{{\cal C}({\cal K},\Gamma)^+}=\pi_P^+\oplus
\pi_P^-.
\end{equation}
Given two basis projections $P_1,P_2$, then
\begin{equation}
\pi_{P_1}^\pm \cong \pi_{P_2}^\pm
\end{equation}
if and only if $P_1-P_2\in{\cal J}_2({\cal K})$ and
{\rm ind}$(P_1,P_2)=+1$, and
\begin{equation}
\pi_{P_1}^\pm \cong \pi_{P_2}^\mp
\end{equation}
if and only if $P_1-P_2\in{\cal J}_2({\cal K})$ and
{\rm ind}$(P_1,P_2)=-1$.
\end{theorem}
On the other hand, it was proven in \cite{Decom} that
pseudo Fock representations $\pi_+$ and $\pi_-$ of
Theorem \ref{evenodd}, when restricted to the even
algebra, remain irreducible but become equivalent.
Summarizing we obtain
\begin{theorem}
\label{restodd}
With notations of Theorem \ref{evenodd}, a representation
$\pi_P\circ\varrho_V$ restricts as follows to the even
algebra ${\cal C}({\cal K},\Gamma)^+$: If $M_V$ is even
we have
\begin{equation}
\pi_P\circ\varrho_V|_{{\cal C}({\cal K},\Gamma)^+}
\cong 2^\frac{M_V}{2}(\pi_{P'}^+\oplus\pi_{P'}^-)
\end{equation}
with $\pi_{P'}^+$ and $\pi_{P'}^-$ mutually
inequivalent and irreducible. If $M_V$ is odd, then
\begin{equation}
\pi_P\circ\varrho_V|_{{\cal C}({\cal K},\Gamma)^+}
\cong 2^\frac{M_V+1}{2} \pi
\end{equation}
with $\pi$ irreducible.
\end{theorem}

\subsection{Restriction to the Algebra of the
Punctured Circle}
Let us consider the algebra of the punctured
circle ${\cal A}(I_\zeta)$. There is no $Y$
in ${\cal A}(I_\zeta)$ and the generators
are of the form
\[ B_{I_\zeta}(f,g) = \hat{\psi}(\hat{f})
\hat{\psi}(\hat{g}), \qquad \hat{f}=f\oplus f,
\quad \hat{g}=g\oplus g. \]
Thus we identify ${\cal A}(I_\zeta)$ as the even
algebra ${\cal C}(L^2(S^1),\Gamma)^+$ and we are
allowed to denote the generators by
\[ B_I(f,g)=\psi(f)\psi(g),\qquad f,g\in L^2(I),
\quad I\subset I_\zeta, \]
i.e.~we work with common CAR algebras. By
construction (\ref{C1}) and (\ref{C2}) it is easy to
see that our representations $\pi_J$, being
non-faithful on ${\cal A}_{\rm univ}^{C^*}$,
fulfill
\[ \pi_J({\cal A}(I_\zeta))=
\pi_J({\cal A}_{\rm univ}^{C^*}),\qquad J=0,
\frac{1}{2},1, \]
the representation theories of ${\cal A}(I_\zeta)$
and ${\cal A}_{\rm univ}^{C^*}$ are obviously the
same. Since ${\cal A}(I_\zeta)\cong
{\cal C}(L^2(S^1),\Gamma)^+$ we can identify
representations $\pi_{\rm NS}$ and $\pi_{\rm R}$ with
GNS representations of quasifree states of
${\cal C}(L^2(S^1),\Gamma)$ and, correspondingly,
representations $\pi_0,\pi_{1/2},\pi_1$
with associated restrictions to the even
subalgebra. This works as follows. Consider
$S_{\rm NS},S_{\rm R}\in{\cal Q}(L^2(S^1),\Gamma)$, the
Neveu Schwarz operator
\begin{equation}
S_{\rm NS}=\sum_{r\in \mathbb{N}_0+\frac{1}{2}} |e_{-r}
\rangle \langle e_{-r} |
\end{equation}
is a basis projection, the Ramond operator
\begin{equation}
S_{\rm R}=\frac{1}{2} |e_0\rangle\langle e_0| +
\sum_{n\in \mathbb{N}} |e_{-n} \rangle\langle e_{-n}|
\end{equation}
is not. By
$({\cal H}_{S_{\rm NS}},\pi_{S_{\rm NS}},|\Omega_{S_{\rm NS}}\rangle)$
and $({\cal H}_{S_{\rm R}},\pi_{S_{\rm R}},|\Omega_{S_{\rm R}}\rangle)$
we denote the GNS triples of the corresponding
quasifree states $\omega_{S_{\rm NS}}$ and $\omega_{S_{\rm R}}$,
respectively. We have
\[ \omega_{S_{\rm NS}} (\psi(e_r)^*\psi(e_r)) =
\langle e_r, S_{\rm NS} e_r \rangle = 0, \qquad r \in
\mathbb{N}_0+\frac{1}{2} \]
and therefore, corresponding to Eq.~(\ref{defNS}),
\[ \pi_{S_{\rm NS}}(\psi(e_r))|\Omega_{S_{\rm NS}} \rangle =0,
\qquad r \in \mathbb{N}_0 + \frac{1}{2}, \]
as well as
\[ \omega_{S_{\rm R}} (\psi(e_n)^*\psi(e_n)) = \langle e_n,
S_{\rm R} e_n \rangle =0, \qquad n \in \mathbb{N} \]
and therefore, corresponding to Eq.~(\ref{defR}),
\[ \pi_{S_{\rm R}}(\psi(e_n))|\Omega_{S_{\rm R}}\rangle=0,
\qquad n\in\mathbb{N}. \]
Since $\omega_{S_{\rm R}}$ is quasifree we obtain
$\omega_{S_{\rm R}}(\psi(e_0))=0$ as the correspondence
to the additional requirement (\ref{ortho}).
Consider an arbitrary element $x=x_{\rm NS}\oplus x_R$
of {\sf Maj}, $x_{\rm NS},x_R\in {\cal C}(L^2(S^1),\Gamma)$.
By (\ref{defNS}), (\ref{defR}) and (\ref{def0}),
and taking into
consideration that $\pi_{\rm NS}$ and $\pi_{\rm R}$ are
defined as cyclic representations of {\sf Maj}, we
identify
\[ \pi_{\rm NS}(x)=\pi_{S_{\rm NS}}(x_{\rm NS}),\qquad
\pi_{\rm R}(x)=\pi_{S_{\rm R}}(x_R). \]
Now ${\cal A}(I_\zeta)$ is generated by bilinear
expressions of {\sf Maj} with $x_{\rm NS}=x_R$.
Therefore, with identification of
${\cal A}(I_\zeta)$ and ${\cal C}(L^2(S^1),\Gamma)^+$,
we have to identify $\pi_{\rm NS}$ with $\pi_{S_{\rm NS}}$,
$\pi_{\rm R}$ with $\pi_{S_{\rm R}}$, and with notations of
Theorem \ref{resteven}, $\pi_0$ with $\pi_{S_{\rm NS}}^+$
and $\pi_1$ with $\pi_{S_{\rm NS}}^-$. Consider our
isometry $V_{1/2}$ (the Bogoliubov operator defined
at the end of subsection 2.2): The kernel of its adjoint
$V_{1/2}^*$ is one-dimensional, spanned
by the vector $2^{-\frac{1}{2}}(e_\frac{1}{2}+
e_{-\frac{1}{2}})$, i.e.~$M_{V_{1/2}}=1$.
By Theorem \ref{evenodd} we find $\pi_{S_{\rm NS}}\circ
\varrho_{V_{1/2}}\cong\pi_+\oplus\pi_-$
with inequivalent, irreducible pseudo Fock
representations $\pi_\pm$, becoming equivalent in
the restriction to the even algebra by Theorem
\ref{restodd}. Since $S_{\rm R}=V_{1/2}^*S_{\rm NS}V_{1/2}$ the
states $\omega_{S_{\rm R}}$ and $\omega_{S_{\rm NS}}\circ
\varrho_{V_{1/2}}$ coincide.
By Eq. (\ref{cyclics}),
$\pi_{S_{\rm NS}}\circ\varrho_{V_{1/2}}$
is indeed a GNS representation of $\omega_{\rm R}$, the
Fock vacuum $|\Omega_{S_{\rm NS}}\rangle$ is cyclic for
$\pi_{S_{\rm NS}}\circ\varrho_{V_{1/2}}$ since
$N_V=0$. This establishes
$\pi_{S_{\rm R}}\cong\pi_{S_{\rm NS}}\circ
\varrho_{V_{1/2}}\cong\pi_+\oplus\pi_-$.
We conclude that the equivalent restrictions
of $\pi_+$ and $\pi_-$ to the even algebra
correspond to the representations $\pi_{1/2}$ and
$\pi_{1/2}'$.

\subsection{Examples of Localized Endomorphisms}
We have seen that, when working on the punctured
circle, one has to deal with even CAR algebras.
Thus we are allowed to define endomorphisms
of ${\cal A}(I_\zeta)$ simply as Bogoliubov
endomorphisms of the underlying algebra
${\cal C}(L^2(S^1),\Gamma)$. We remark that our
endomorphisms $\varrho_{1/2}$ and $\varrho_1$ of
${\cal A}_{\rm univ}^{C^*}$ do not restrict to
endomorphisms of ${\cal A}(I_\zeta)$ because of the
different entries $V_J$ and $V_J'$ in matrices
$\hat{V}_J$, $J=\frac{1}{2},1$ of Eq.~(\ref{matrix}).
For constructing localized endomorphisms,
we admit as localization
regions all open intervals such that their
closure is contained in $I_\zeta$, i.e.~elements
of the set
\begin{equation}
\label{Izeta}
{\cal J}_\zeta = \{ I\in {\cal J}\,\,|\,\,
\zeta\in I' \}.
\end{equation}
($I'$ always denotes the interior of the
complement $I^c=S^1\setminus I$.)
As usual, we define an endomorphism $\varrho$ of
${\cal A}(I_\zeta)$ to be localized in some
interval $I\in{\cal J}_\zeta$ if $\varrho(A)=A$
for all $A\in{\cal A}(I_1)$, $I_1\in{\cal J}_\zeta,
I_1\cap I=\emptyset$. We present localized
endomorphisms as Bogoliubov endomorphisms which
are induced by pseudolocalized isometries.
A Bogoliubov operator
$V\in{\cal I}(L^2(S^1),\Gamma)$ is called
pseudolocalized in $I\in{\cal J}_\zeta$ if for all
$f\in L^2(S^1)$
\[ (Vf)(z)=\sigma_\pm f(z),\qquad z\in I_\pm,
\qquad \sigma_\pm \in \{-1,1\}, \]
where $I_+,I_-$ denote the two connected components
of $I'\cap I_\zeta$. Moreover, $V$ is called even,
if $\sigma_+=\sigma_-$, and odd, if $\sigma_+=
-\sigma_-$. Clearly, a pseudolocalized Bogoliubov
operator induces a localized endomorphism of the
even algebra ${\cal A}(I_\zeta)=
{\cal C}(L^2(S^1),\Gamma)^+$. We give the following
examples
\begin{definition}
\label{rho1loc}
Let $h\in L^2(S^1)$ be a real
(i.e.~$\Gamma$-invariant) function, $\|h\|=1$ and
{\rm supp}$(h)\subset I$ for some $I\in{\cal J}_\zeta$.
Define $W\in{\cal I}(L^2(S^1),\Gamma)$,
\begin{equation}
W= 2|h\rangle\langle h|-{\bf 1}
\end{equation}
and the automorphism $\varrho_1^{\rm loc}=\varrho_W$
of ${\cal C}(L^2(S^1),\Gamma)$.
\end{definition}
Obviously, $W$ is even pseudolocalized and
$\varrho_1^{\rm loc}$ therefore, when restricted to
${\cal A}(I_\zeta)$, localized in $I$.
\begin{lemma}
The automorphism $\varrho_1^{\rm loc}$ is inner in
${\cal C}(L^2(S^1),\Gamma)$. In restriction to the
even algebra ${\cal A}(I_\zeta)$ it leads to
\begin{equation}
\pi_0\circ\varrho_1^{\rm loc} \cong \pi_1.
\end{equation}
\end{lemma}
{\it Proof.} One easily checks that $\varrho_W$ is
implemented in ${\cal C}(L^2(S^1),\Gamma)$ by the
unitary self-adjoint $q(W)=\sqrt{2}\psi(h)$, for all
$f\in L^2(S^1)$ we have
\begin{eqnarray*}
q(W)\psi(f)q(W)
&=& 2\psi(h)\psi(f)\psi(h) \\
&=& 2\{\psi(h),\psi(f)\}\psi(h) - 2\psi(f)\psi(h)\psi(h) \\
&=& 2\langle h,f \rangle \psi(h)  - \psi(f) \\
&=& \psi( 2\langle h, f \rangle h - f)  \\
&=& \psi(Wf).
\end{eqnarray*}
Thus $\varrho_W$ is implemented in $\pi_{S_{\rm NS}}$ by
$Q(W)=\pi_{S_{\rm NS}}(q(W))$. Obviously we have
$\bar{\alpha}_{-1}(Q(U))=-Q(U)$ and $S_{\rm NS}-W^*S_{\rm NS}W$
is Hilbert Schmidt class. By Proposition \ref{ind} and
Theorem \ref{resteven} we conclude
\[ \pi_{S_{\rm NS}}^+\circ\varrho_W \cong \pi_{S_{\rm NS}}^- \]
which proves the lemma, q.e.d.

In the following we are searching for a localized
endomorphism $\varrho_{1/2}^{\rm loc}$ which leads
to a representation being equivalent to $\pi_{1/2}$.
It turns out that the discussion becomes much more
complicated. First we fix our point $\zeta$ to be
$\zeta=-1$, without loss of generality. Further, we
choose the localization region $I$ to be $I_2$,
\[ I_2 = \left\{ z=\E^{\I\phi} \in S^1 \left|
-\frac{\pi}{2} <
\phi < \frac{\pi}{2} \right. \right\} \]
so that the open complement $I_2'$ is divided by
$\zeta$ into $I_-$ and $I_+$,
\begin{eqnarray*}
I_- &=& \left\{ z=\E^{\I\phi} \in S^1 \left| -\pi <
\phi < -\frac{\pi}{2} \right. \right\}, \\
I_+ &=& \left\{ z=\E^{\I\phi} \in S^1 \left|
\frac{\pi}{2} < \phi < \pi \right. \right\}.
\end{eqnarray*}
The Hilbert space $L^2(S^1)$ decomposes into a direct sum,
\[ L^2(S^1) = L^2(I_-) \oplus L^2(I_2) \oplus L^2(I_+).\]
By $P_{I_+}$, $P_{I_-}$ we denote the projections on the
subspaces $L^2(I_+)$, $L^2(I_-)$, respectively.
Define functions on $S^1$ by
\[ e_a^{(2)}(z)= \left\{ \begin{array}{cl}
\sqrt{2}z^{2a} & \qquad z\in I_2 \\
0 & \qquad z\in S^1\setminus I_2 \end{array} \right.,
\qquad a\in \frac{1}{2}\mathbb{Z}. \]
With
\[ \left\{e_r^{(2)}, r\in \mathbb{Z}+\frac{1}{2}
\right\},\qquad
\{ e_n^{(2)}, n\in\mathbb{Z} \} \]
we then obtain two orthonormal bases of the subspace
$L^2(I_2)\subset L^2(S^1)$.
\begin{definition}
\label{def1/2loc}
We define\footnote{The definition of $V'$ was already
suggested by Mack and Schomerus \cite{MS1}.}
Bogoliubov operators $V,V'\in {\cal I}(L^2(S^1),\Gamma)$
as follows,
\begin{eqnarray}
V &=& P_{I_-}-P_{I_+}+ \frac{\I}{\sqrt{2}} \Big(
|e_\frac{1}{2}^{(2)}\rangle\langle e_0^{(2)}|-
|e_{-\frac{1}{2}}^{(2)}\rangle\langle e_0^{(2)}|
\Big) \\
&& \qquad + \I \sum_{n=1}^\infty\Big(
|e_{n+\frac{1}{2}}^{(2)}\rangle
\langle e_n^{(2)} | - |e_{-n-\frac{1}{2}}^{(2)}
\rangle\langle e_{-n}^{(2)}|\Big),\nonumber \\
V' &=& P_{I_-}-P_{I_+}+ \I \sum_{n=1}^\infty\Big(
|e_n^{(2)}\rangle\langle e_{n-\frac{1}{2}}^{(2)} |
-|e_{-n}^{(2)}\rangle\langle
e_{-n+\frac{1}{2}}^{(2)}|\Big).
\end{eqnarray}
and let $\varrho_{1/2}^{\rm loc}$ and
$\sigma_{1/2}^{\rm loc}$ be the endomorphisms of
${\cal C}(L^2(S^1),\Gamma)$ defined by
$\varrho_{1/2}^{\rm loc}=\varrho_V$ and
$\sigma_{1/2}^{\rm loc}=\varrho_{V'}$.
\end{definition}
Obviously, $V$ and $V'$ are odd pseudolocalized and
$\varrho_{1/2}^{\rm loc}$ and $\sigma_{1/2}^{\rm loc}$
therefore, when restricted to ${\cal A}(I_\zeta)$,
localized in $I_2$.

\subsection{Analysis of $\varrho_{1/2}^{\rm loc}$
and $\sigma_{1/2}^{\rm loc}$}
In this subsection we establish that
$\pi_0\circ\varrho_{1/2}^{\rm loc}$ is unitarily
equivalent to $\pi_{1/2}$. Furthermore we identify
the unitary equivalence class of
$\pi_0\circ\varrho_{1/2}^{\rm loc}\varrho_{1/2}^{\rm loc}$.
The first step is the following
\begin{lemma}
\label{HS}
The following operators are Hilbert Schmidt class,
\begin{eqnarray}
\label{HS1}
V^* S_{\rm NS} V - S_{\rm R} &\in& {\cal J}_2(L^2(S^1)),\\
\label{HS2}
V S_{\rm NS} V^* - S_{\rm R} &\in& {\cal J}_2(L^2(S^1)),\\
{V'}^* S_{\rm NS} V' - S_{\rm R} &\in& {\cal J}_2(L^2(S^1)),\\
V' S_{\rm NS} {V'}^* - S_{\rm R} &\in& {\cal J}_2(L^2(S^1)).
\end{eqnarray}
\end{lemma}
Because the proof is ugly work it is banished to the
appendix. For drawing our first conclusions of Lemma
\ref{HS}, we remember an estimate which was given by
Powers and St{\o}rmer \cite{Pow}: For positive operators
$A,B\in{\cal B}({\cal K})$ the following inequality holds:
\begin{equation}
\label{normest}
\| A^\frac{1}{2}-B^\frac{1}{2} \|_2^2 \le \| A-B \|_1,
\end{equation}
where for $T\in{\cal B}(L^2(S^1))$ by $\|T\|_1$ is denoted
the trace norm
\[ \|T\|_1 = \mbox{tr}\left((T^*T)^\frac{1}{2}\right),\]
and by $\|T\|_2$ the Hilbert Schmidt norm
\[ \| T \|_2 = ( \mbox{tr} (T^*T) )^\frac{1}{2}. \]
This estimate will be used to see that the operators
$V^*S_{\rm NS}V$ and ${V'}^*S_{\rm NS}V'$ differ from their
positive square roots only by Hilbert Schmidt operators,
\begin{eqnarray*}
\| (V^*S_{\rm NS}V)^\frac{1}{2} - V^*S_{\rm NS}V \|_2^2 &\le&
\| V^*S_{\rm NS}V - (V^*S_{\rm NS}V)^2 \|_1 \\ &=&
\| V^*S_{\rm NS}({\bf 1}-VV^*)S_{\rm NS}V \|_1 \\
&\le& \| V \|^2 \|S_{\rm NS}\|^2 \|{\bf 1}-VV^*\|_1.
\end{eqnarray*}
Since ${\bf 1}-VV^*$ is a rank one projection and
therefore trace class, the right hand side is finite.
Obviously, the same calculation runs for $V'$.
More easily one finds
\[ \|S_{\rm R}^\frac{1}{2}-S_{\rm R}\|_2 = \left\| \left(
\frac{1}{\sqrt{2}}-\frac{1}{2}\right) |e_0\rangle\langle
e_0| \right\|_2 = \frac{1}{\sqrt{2}}-\frac{1}{2}.\]
It follows immediately from Lemma \ref{HS}:
\[ (V^*S_{\rm NS}V)^\frac{1}{2}-S_{\rm R}^\frac{1}{2} \in
{\cal J}_2(L^2(S^1)), \qquad
({V'}^*S_{\rm NS}V')^\frac{1}{2}-S_{\rm R}^\frac{1}{2} \in
{\cal J}_2(L^2(S^1)). \]
Applying Theorem \ref{Araki} this yields
$\omega_{V^*S_{\rm NS}V}\approx\omega_{\rm R}$ and thus
\[ \pi_{S_{\rm NS}} \circ \varrho_V \approx \pi_{S_{\rm R}},\]
the same holds for $\varrho_{V'}$.
We have already discussed that $\pi_{S_{\rm R}}$
decomposes into two inequivalent pseudo Fock
representations. Using Theorem \ref{evenodd}, the
same is true for $\pi_{S_{\rm NS}}\circ\varrho_V$ and
$\pi_{S_{\rm NS}}\circ\varrho_{V'}$ since
$M_V=M_{V'}=1$. Thus we have
\[ \pi_{S_{\rm NS}}\circ\varrho_V \cong \pi_+ \oplus
\pi_- \cong \pi_{S_{\rm NS}}\circ\varrho_{V'}. \]
In restriction to the even algebra
${\cal C}(L^2(S^1),\Gamma)^+$, the representations
$\pi_+$ and $\pi_-$ become equivalent and have
to be identified with $\pi_{1/2}$. This means
$\pi_{S_{\rm NS}}^+\circ\varrho_V\cong
\pi_{S_{\rm NS}}^-\circ\varrho_V$ and
$\pi_{S_{\rm NS}}^+\circ\varrho_{V'}\cong
\pi_{S_{\rm NS}}^-\circ\varrho_{V'}$. We have
proven
\begin{theorem}
\label{1/2loc}
The representations of ${\cal A}(I_\zeta)$ obey
\begin{eqnarray}
\pi_0\circ\varrho_{1/2}^{\rm loc} \cong
\pi_1\circ\varrho_{1/2}^{\rm loc} &\cong& \pi_{1/2}\\
\pi_0\circ\sigma_{1/2}^{\rm loc} \cong
\pi_1\circ\sigma_{1/2}^{\rm loc} &\cong& \pi_{1/2}.
\end{eqnarray}
\end{theorem}
Let us now consider the squares $\varrho_V^2=\varrho_{V^2}$
and $\varrho_{V'}^2=\varrho_{{V'}^2}$.
\begin{proposition}
\label{square}
The representations of ${\cal A}(I_\zeta)$ obey
\begin{eqnarray}
\pi_0\circ\varrho_{1/2}^{\rm loc}\varrho_{1/2}^{\rm loc}
&\cong& \pi_0 \oplus \pi_1 \\
\pi_0\circ\sigma_{1/2}^{\rm loc}\sigma_{1/2}^{\rm loc}
&\cong& \pi_0 \oplus \pi_1.
\end{eqnarray}
\end{proposition}
{\it Proof.} If we multiply the operator in relation
(\ref{HS2}) with $V^*$ from the left and with $V$ from the
right we get
\[ V^*S_{\rm R}V-S_{\rm NS}\in{\cal J}_2(L^2(S^1)). \]
Since relation (\ref{HS1}) holds we can replace $S_{\rm R}$
by $V^*S_{\rm NS}V$, this yields
\[ V^*V^*S_{\rm NS}VV-S_{\rm NS} \in
{\cal J}_2(L^2(S^1)). \]
In the same way one obtains
\[ {V'}^*{V'}^*S_{\rm NS}V'V'-S_{\rm NS} \in
{\cal J}_2(L^2(S^1)). \]
Now the operators ${\bf 1}-(VV)(VV)^*$ and
${\bf 1}-(V'V')(V'V')^*$ are  rank two projections,
so that we can again conclude
\[ (V^*V^*S_{\rm NS}VV)^\frac{1}{2}-S_{\rm NS} \in
{\cal J}_2(L^2(S^1)). \]
and
\[ ({V'}^*{V'}^*S_{\rm NS}V'V')^\frac{1}{2}-S_{\rm NS} \in
{\cal J}_2(L^2(S^1)). \]
By using Theorem \ref{Araki} we obtain for the states
\[ \omega_{S_{\rm NS}} \circ \varrho_V^2
\approx \omega_{S_{\rm NS}}, \qquad
\omega_{S_{\rm NS}} \circ \varrho_{V'}^2
\approx \omega_{S_{\rm NS}}. \]
Moreover, the kernel of $V^*$ is spanned by the
$\Gamma$-invariant, normed vector
\[ f_0^{(2)} = \frac{1}{\sqrt{2}}(e_\frac{1}{2}^{(2)}
+ e_{-\frac{1}{2}}^{(2)}).\]
and the kernel of ${V'}^*$ is spanned by $e_0^{(2)}$.
Thus, ker$(V^*)^2$ (resp.~ker$({V'}^*)^2$) is spanned
by orthonormal vectors $f_0^{(2)}$ and $Vf_0^{(2)}$
(resp.~$e_0^{(2)}$ and $V'e_0^{(2)}$)
i.e.~$M_{V^2}=M_{{V'}^2}=2$; we conclude
\[ \pi_{S_{\rm NS}}\circ\varrho_V^2 \cong \pi_{S_{\rm NS}}
\oplus \pi_{S_{\rm NS}} \cong \pi_{S_{\rm NS}}\circ\varrho_{V'}^2 \]
by Theorem \ref{evenodd}. In restriction to the even algebra,
identified with ${\cal A}(I_\zeta)$, this reads
\begin{eqnarray*}
\pi_0 \circ \varrho_{1/2}^{\rm loc}
\varrho_{1/2}^{\rm loc} \oplus
\pi_1 \circ \varrho_{1/2}^{\rm loc}
\varrho_{1/2}^{\rm loc} &\cong&
\pi_0 \oplus \pi_1 \oplus \pi_0 \oplus \pi_1 ,\\
\pi_0 \circ \sigma_{1/2}^{\rm loc}
\sigma_{1/2}^{\rm loc} \oplus
\pi_1 \circ \sigma_{1/2}^{\rm loc}
\sigma_{1/2}^{\rm loc} &\cong&
\pi_0 \oplus \pi_1 \oplus \pi_0 \oplus \pi_1.
\end{eqnarray*}
We have to assign the irreducible representations on
the right hand side to the representations on the
left. By Theorem \ref{1/2loc} we find
$\pi_0\circ\varrho_{1/2}^{\rm loc}\cong
\pi_1\circ\varrho_{1/2}^{\rm loc}$ and therefore
$\pi_0\circ\varrho_{1/2}^{\rm loc}\varrho_{1/2}^{\rm loc}
\cong\pi_1\circ\varrho_{1/2}^{\rm loc}
\varrho_{1/2}^{\rm loc}$. Using the same argument,
one obtains
$\pi_0\circ\sigma_{1/2}^{\rm loc}\sigma_{1/2}^{\rm loc}
\cong\pi_1\circ\sigma_{1/2}^{\rm loc}
\sigma_{1/2}^{\rm loc}$, q.e.d.

\section{Extension to von Neumann Algebras}
In the DHR theory one usually works with
local von Neumann algebras instead of local
$C^*$-algebras. This formalism allows to
discover intertwiners in the observable
algebra and this is crucial for the analysis
of statistics and fusion. It is our aim to find
a description of the chiral Ising model as
close as possible to the DHR formalism. Therefore
we have to extend our local $C^*$-algebras
${\cal A}(I)$,  to their weak closures in the
vacuum representation.

\subsection{The Net of Local von Neumann Algebras}
For intervals $I\in {\cal J}$ with non-empty open
complement, we define local von Neumann algebras
\[ {\cal R}(I) = \pi_0({\cal A}(I))'', \qquad
I\in {\cal J},\quad I'\neq \emptyset. \]
By M{\"o}bius covariance (some details are
presented in the appendix), this
defines a so-called covariant precosheaf on the
circle. In particular, we have Haag duality on
the circle \cite{BGL,Buch},
\begin{equation}
\label{HD}
{\cal R}(I)'={\cal R}(I').
\end{equation}
Since the set of intervals
$\{I\in{\cal J}$,$I'\neq\emptyset\}$
is not directed we cannot define a
global algebra as the $C^*$-norm closure of the union
of all local algebras. Following Fredenhagen, Rehren and
Schroer \cite{FRS2}, one could instead introduce the
corresponding universal algebra. But in our model it
seems to be much more comfortable to define a
quasilocal algebra of the punctured circle. (The set
${\cal J}_\zeta$, Eq. (\ref{Izeta}), is directed.)
This works as follows.
We fix again our ``point at infinity'', without loss of
generality, to be $\zeta=-1$ and admit only intervals
$I\in{\cal J}_\zeta$. Then we define our algebra
$\mathfrak{A}_\zeta$ of quasilocal observables to be the
norm closure of all such local von Neumann
algebras,
\begin{equation}
\label{Azeta}
\mathfrak{A}_\zeta = \overline{\bigcup_{I\in{\cal J}_\zeta}
{\cal R}(I)}.
\end{equation}
Now choose some interval $I\in{\cal J}_\zeta$. Let us
denote the von Neumann algebra generated by all
${\cal R}(I_0)$, $I_0\in{\cal J}_\zeta$,
$I_0\cap I=\emptyset$ by ${\cal R}_\zeta(I')$.
Obviously we have
\[ {\cal R}_\zeta(I') \subset {\cal R}(I'). \]
We claim that equality holds, that means Haag duality
holds also on the punctured circle.
\begin{lemma}
We have Haag duality on the punctured circle.
For $I\in{\cal J}_\zeta$ the following relation holds,
\begin{equation}
\label{HDpunc}
{\cal R}(I)'={\cal R}_\zeta(I').
\end{equation}
\end{lemma}
{\it Proof.} We have to show
${\cal R}_\zeta(I')={\cal R}(I')$. It is sufficient to
show that each generator $\pi_0(\psi(f)\psi(g))$,
supp$(f)$,supp$(g)\subset I'$ of ${\cal R}(I')$ is
a weak limit point of a net constructed out of
elements in ${\cal R}_\zeta(I')$.
Let $I_0\in{\cal J}_\zeta$,
$I_0\supset I$ be an interval such that $I_0'$ is
a small neighborhood of $\zeta$. Let $\chi_0$ be the
characteristic function of $I_0$ and
$f_0=\chi_0f$, $g_0=\chi_0g$. Since
$\pi_0(\psi(f_0)\psi(g_0))$ converges in $C^*$-norm
to $\pi_0(\psi(f)\psi(g))$ by inequality (\ref{Cstern}) if
$I_0'$ shrinks to the point $\zeta$ it suffices to show
that $\pi_0(\psi(f_0)\psi(g_0))$ is such a limit
point for all such $I_0$. Now let us denote the two
connected components of $I'\setminus\{\zeta\}$ by
$I_+$ and $I_-$. Define $\chi_\pm$ to be the
characteristic functions of $I_\pm$, and we write
$f_\pm=\chi_\pm f_0$, $g_\pm=\chi_\pm g_0$. Then
\begin{eqnarray*}
\pi_0(\psi(f_0)\psi(g_0)) &=& \pi_0(\psi(f_+)\psi(g_+))
+ \pi_0(\psi(f_-)\psi(g_-)) \\ && \qquad
+ \pi_0(\psi(f_+)\psi(g_-)) + \pi_0(\psi(f_-)\psi(g_+))
\end{eqnarray*}
Clearly, the first two terms on the
right hand side are elements of ${\cal R}_\zeta(I')$.
We show that the third (then, by
symmetry, also the fourth) is a weak limit point
described as above. Choose sequences
$\{h_\pm^{(n)},n\in\mathbb{N}\}$, where
$h_\pm^{(n)}\in L^2(S^1)$ are functions,
$\|h_\pm^{(n)}\|=1$, supp$(h_\pm^{(n)})\subset
I_\pm^{(n)}$, and $I_\pm^{(n)}\in{\cal J}_\zeta$
are intervals, $I_\pm^{(n)}\subset I_\pm\cap I_0'$,
shrinking to the point $\zeta$. Then define for
$n\in\mathbb{N}$
\[ Z_n=\pi_0(\psi(h_+^{(n)})\psi(h_-^{(n)})) \in
{\cal R}(I'). \]
By M{\"o}bius covariance of the vacuum sector (appendix),
we can choose the $h_\pm^{(n)}$ such that the $Z_n$
are related by M{\"o}bius transformations (dilations).
For all $n$ we have $\|Z_n\|\le 1$. It follows that
there exists a weakly convergent subnet
$\{X_\alpha,\alpha\in\iota\}$, that means there is
a function $F:\iota\rightarrow\mathbb{N}$ ($\iota$ a
directed set) with the property that
$X_\alpha=Z_{F(\alpha)}$ for all $\alpha\in\iota$,
and that for each $n'\in\mathbb{N}$ there is an
$\alpha'\in\iota$ such that $\alpha\succ\alpha'$
implies $F(\alpha)\ge n'$ \cite{RS1}. The weak limit
point of the net $\{X_\alpha,\alpha\in\iota\}$ in
${\cal R}(I')$ will be denoted by $X$,
\[ \mbox{w-}\lim_\alpha X_\alpha = X. \]
For each $I_1\in {\cal J}_\zeta$ all elements
$R\in{\cal R}(I_1)$ commute with $Z_n$ for
sufficiently large $n$, thus $[X,R]=0$. It follows
\[ [X,A]=0, \qquad A\in\mathfrak{A}_\zeta, \]
and, by irreducibility of the vacuum representation,
$X$ is a complex number, $X=\lambda {\bf 1}$. We have
\[ \lambda = \langle\Omega_0|X|\Omega_0\rangle =
\lim_\alpha \langle\Omega_0|X_\alpha|\Omega_0\rangle
= \lim_\alpha \langle\Omega_0|Z_{F(\alpha)}|\Omega_0\rangle
= \langle\Omega_0|Z_1|\Omega_0\rangle \]
by M{\"o}bius invariance of the vacuum.
We claim that we can choose $h_+^{(1)}$ and $h_-^{(1)}$
such that $\lambda\neq 0$. Recall the definition of the
Hardy space
\[ H^2 = \{ f\in L^2(S^1)\,\,|\,\, \langle e_{-n},f \rangle
=0, \,\, n=1,2,\ldots \}. \]
A Theorem of F. and M. Riesz (see e.g.~\cite{Doug}, Th.~6.13)
states that
\begin{equation}
\label{Riesz}
f\in H^2,f\neq 0 \qquad \Longrightarrow \qquad
f(z)\neq 0 \mbox{ almost everywhere}.
\end{equation}
For example, for a given non-zero function
$k\in S_{\rm NS}L^2(S^1)$
we find that $k'\in H^2$ where $k'(z)=z^\frac{1}{2}
\overline{k(z)}$. So $k'$ and hence $k$ cannot vanish
on a set of non-zero measure. Now we have
\[ \lambda = \langle\Omega_0|Z_1|\Omega_0\rangle =
\omega_0(\psi(h_+^{(1)})\psi(h_-^{(1)}))=
\langle\Gamma h_+^{(1)},S_{\rm NS} h_-^{(1)} \rangle. \]
Define $k=S_{\rm NS}h_-^{(1)}$ for a given $h_-^{(1)}$
as above. We find $k\neq 0$, otherwise
$\Gamma h_-^{(1)}\in S_{\rm NS}L^2(S^1)$ in
contradiction to the fact that $h_-^{(1)}$ and
hence $\Gamma h_-^{(1)}$ vanishes outside
$I_-^{(1)}$. It follows $k(z)\neq 0$ almost everywhere
and hence
\[ c=\int_{I_+^{(1)}} |k(z)|^2 \frac{dz}{2\pi\I z}
\neq 0 \]
If we define, for instance,
$h_+^{(1)}(z)=c^{-\frac{1}{2}} \chi_{I_+^{(1)}}(z)
\overline{k(z)}$
where $\chi_{I_+^{(1)}}$ denotes the characteristic
function of $I_+^{(1)}$ then indeed
$\lambda=c^\frac{1}{2}\neq 0$. So we can compute
\begin{eqnarray*}
\pi_0(\psi(f_+)\psi(g_-)) &=& \lambda^{-1} \pi_0
(\psi(f_+)\psi(g_-)) X \\
&=& \lambda^{-1} \mbox{w-}\lim_\alpha
\pi_0(\psi(f_+)\psi(g_-))X_\alpha \\
&=& \lambda^{-1} \mbox{w-}\lim_\alpha
\pi_0(\psi(f_+)\psi(g_-))Z_{F(\alpha)} \\
&=& \lambda^{-1} \mbox{w-}\lim_\alpha
\pi_0(\psi(f_+)\psi(g_-)\psi(h_+^{(F(\alpha))})
\psi(h_-^{(F(\alpha))}))\\
&=& \lambda^{-1} \mbox{w-}\lim_\alpha
\pi_0(\psi(f_+)\psi(h_+^{(F(\alpha))})
\psi(h_-^{(F(\alpha))})\psi(g_-))\\
&=& \lambda^{-1} \mbox{w-}\lim_\alpha
\pi_0(\psi(f_+)\psi(h_+^{(F(\alpha))}))
\pi_0(\psi(h_-^{(F(\alpha))})\psi(g_-))
\end{eqnarray*}
i.e.~$\pi_0(\psi(f_+)\psi(g_-))$ is indeed a weak limit
point of a net of elements in ${\cal R}_\zeta(I')$,
q.e.d.

\subsection{Extension}
Because we work on the punctured circle the
vacuum representation is faithful i.e.~$\pi_0$
acts faithfully on ${\cal A}(I_\zeta)$. Thus we
are allowed to use the common convention of
identifying observables $A$ with their vacuum
representers $\pi_0(A)$. Passing over to von
Neumann algebras, we consider the vacuum
representation acting as the identity on
${\cal R}(I)$, and in the same fashion, we treat
local $C^*$-algebras ${\cal A}(I)$ as subalgebras
of ${\cal R}(I)$, $I\in{\cal J}_\zeta$. Now
we have to check whether we can canonically
extend our representations $\pi_J$ and
endomorphisms $\varrho_J^{\rm loc}$,
$J=\frac{1}{2},1$, to the von Neumann algebras
${\cal R}(I)$, $I\in{\cal J}_\zeta$, and the
global (quasilocal) $C^*$-algebra $\mathfrak{A}_\zeta$
they generate. Thus we are looking for
isomorphisms
\[ \hat{\pi}_J \,\,: \,\, {\cal A}(I)''=
{\cal R}(I) \longrightarrow \pi_J({\cal A}(I))'' \]
satisfying $\hat{\pi}_J(A)=\pi_J(A)$ if
$A\in{\cal A}(I)$, $I\in{\cal J}_\zeta$,
$J=\frac{1}{2},1$. This means exactly that we have
to check whether the representations $\pi_J$ are
quasiequivalent to the identity (vacuum
representation) on local algebras ${\cal A}(I)$.
\begin{theorem}[Local Normality]
\label{locnorm}
In restriction to local $C^*$-algebras ${\cal A}(I)$,
$I\in{\cal J}_\zeta$, the representations $\pi_J$ are
quasiequivalent to the vacuum representation
$\pi_0=id$,
\begin{equation}
\pi_J|_{{\cal A}(I)} \approx \pi_0|_{{\cal A}(I)},
\qquad I\in{\cal J}_\zeta,\quad J=\frac{1}{2},1.
\end{equation}
\end{theorem}
{\it Proof.} First we consider the case $J=1$. We
have to show that
\[ \pi_{S_{\rm NS}}^-|_{{\cal C}(L^2(I),\Gamma)^+}
\approx \pi_{S_{\rm NS}}^+|_{{\cal C}(L^2(I),\Gamma)^+}.\]
We have already proven that $\pi_{S_{\rm NS}}^-\cong
\pi_{S_{\rm NS}}^+\circ\varrho_W$ on
${\cal C}(L^2(S^1),\Gamma)^+$. We show that
$\pi_{S_{\rm NS}}^+\circ\varrho_W$ and $\pi_{S_{\rm NS}}^+$,
when restricted to ${\cal C}(L^2(I),\Gamma)^+$,
are unitarily equivalent ($I\in{\cal J}_\zeta$).
In ${\cal C}(L^2(S^1),\Gamma)$, $\varrho_W$ is
implemented by the unitary $q(W)=\sqrt{2}\psi(h)$.
Choose a real function $h'\in L^2(S^1)$ such that
$\|h'\|=1$ and supp$(h')\subset I_0$ for some
$I_0\in{\cal J}_\zeta$, $I_0\cap I=\emptyset$ and
set $U=\sqrt{2}\psi(h')$. Then $q(W)U$ is a unitary
element of ${\cal C}(L^2(S^1),\Gamma)^+$ and for
$x\in{\cal C}(L^2(I),\Gamma)^+$ we have
\[q(W)Ux(q(W)U)^*=q(W)UxUq(W)=q(W)xq(W)=\varrho_W(x)\]
and therefore
\[ \pi_{S_{\rm NS}}^+\circ\varrho_W(x)=\pi_{S_{\rm NS}}^+
(q(W)U)\pi_{S_{\rm NS}}^+(x)\pi_{S_{\rm NS}}^+(q(W)U)^{-1} \]
which proves the statement. Now consider
$J=\frac{1}{2}$. By the following Lemma \ref{locquasi}
we have quasiequivalence of $\pi_{S_{\rm NS}}$ and
$\pi_{S_{\rm R}}$ on ${\cal C}(L^2(I),\Gamma)$. In
restriction to the even subalgebra, the irreducible
$\pi_{S_{\rm NS}}$ splits into
$\pi_{S_{\rm NS}}^+\oplus\pi_{S_{\rm NS}}^-$, and the two
irreducible subrepresentations $\pi_+$ and $\pi_-$
of $\pi_{S_{\rm R}}$ become equivalent to an irreducible
representation $\pi$.
Thus locally one obtains
\[ (\pi_{S_{\rm NS}}^+\oplus\pi_{S_{\rm NS}}^-)
|_{{\cal C}(L^2(I),\Gamma)^+} \approx 2\pi
|_{{\cal C}(L^2(I),\Gamma)^+} \]
where $\pi$ corresponds to the representation
$\pi_{1/2}$. Having already established the local
equivalence of $\pi_{S_{\rm NS}}^+$ and $\pi_{S_{\rm NS}}^-$
this proves the theorem, q.e.d.
\begin{lemma}
\label{locquasi}
For $I\in{\cal J}_\zeta$ we have the local
quasiequivalence
\begin{equation}
\pi_{S_{\rm NS}}|_{{\cal C}(L^2(I),\Gamma)}
\approx \pi_{S_{\rm R}}|_{{\cal C}(L^2(I),\Gamma)}.
\end{equation}
\end{lemma}
{\it Proof.} We first claim that
$|\Omega_{S_{\rm NS}}\rangle$
and $|\Omega_{S_{\rm R}}\rangle$ remain cyclic for
$\pi_{S_{\rm NS}}({\cal C}(L^2(I),\Gamma))$ and
$\pi_{S_{\rm R}}({\cal C}(L^2(I),\Gamma))$, respectively.
Denote by $P_I$ the projection onto
$L^2(I)\subset L^2(S^1)$. Then for $\pi_{S_{\rm NS}}$ the
statement is a consequence (of the arguments in the proof)
of Araki's Lemma 4.8 in \cite{Ara1}, because
\[ ({\bf 1}-P_I) L^2(S^1) \cap S_{\rm NS} L^2(S^1) =
\{0\}. \]
(If $k\in S_{\rm NS}L^2(S^1)$, $k\neq 0$, then
$k'(z)=z^\frac{1}{2}\overline{k(z)}$ is a function
in $H^2$ and hence again by Eq.~(\ref{Riesz})
$k$ and $k'$ cannot vanish in the whole interval $I$.)
An analogous argument runs for $\pi_{S_{\rm R}}$, because
it is a direct sum of inequivalent pseudo Fock
representations $\pi_+$ and $\pi_-$ which restrict
to Fock representations (see \cite{Ara1} for details)
$\pi_{P_{\rm R}}$ of ${\cal C}((P_{\rm R}+\Gamma P_{\rm R}\Gamma)
L^2(S^1),\Gamma)$ each, where
\[ P_{\rm R}=\sum_{n=1}^\infty |e_{-n}\rangle\langle e_{-n}|\]
is a basis projection of
$L_0^2(S^1)=(P_{\rm R}+\Gamma P_{\rm R}\Gamma)L^2(S^1)$. If
$P_I^{(0)}$ denotes the projection onto
$L_0^2(I)=L_0^2(S^1)\cap L^2(I)$ one finds again
\[  ({\bf 1}-P_I^{(0)}) L_0^2(S^1) \cap P_{\rm R} L_0^2(S^1) =
\{0\} \]
because we have
$P_I^{(0)}=P_I-\langle e_0,P_I e_0\rangle^{-1}
P_I|e_0\rangle\langle e_0|P_I$ and thus a function
$k\in({\bf 1}-P_I^{(0)})L_0^2(S^1)$ is constant in $I$,
$k(z)=c$, $z\in I$ and $c\in\mathbb{C}$.
On the other hand, for $k\in P_{\rm R}L_0^2(S^1)$ we find
$\Gamma k - \overline{c}\in H^2$. Now $\Gamma k-\overline{c}$
vanishes in $I$, so it follows by Eq.~(\ref{Riesz})
that $\Gamma k-\overline{c}=0$, i.e.~$k$ is constant on the
whole circle, $k=c$. But we have $\langle e_0,k\rangle=0$ for
$k\in P_{\rm R}L_0^2(S^1)$ and hence $k=0$. So we conclude that
the GNS vector $|\Omega_{P_{\rm R}}\rangle$ remains cyclic for
$\pi_{P_{\rm R}}({\cal C}(L_0^2(I),\Gamma))$, thus
vectors $|\Omega_\pm\rangle\equiv|\Omega_{P_{\rm R}}\rangle$
for $\pi_\pm({\cal C}(L^2(I),\Gamma))$. By inequivalence of
$\pi_+$ and $\pi_-$, the GNS vector
$|\Omega_{S_{\rm R}}\rangle=2^{-\frac{1}{2}}(|\Omega_+\rangle
\oplus|\Omega_-\rangle)$ remains cyclic for
$\pi_{S_{\rm R}}({\cal C}(L^2(I),\Gamma))$. Thus, for proving
the lemma, we have to show that states
$\omega_{S_{\rm NS}}$ and $\omega_{S_{\rm R}}$ are quasiequivalent
on ${\cal C}(L^2(I),\Gamma)$. We have to show
\[ (P_IS_{\rm NS}P_I)^\frac{1}{2} - (P_IS_{\rm R}P_I)^\frac{1}{2}
\in {\cal J}_2(L^2(S^1)). \]
Using the inequality (\ref{normest}) it is
sufficient to show
\[ \|P_IS_{\rm NS}P_I-P_IS_{\rm R}P_I\|_1 < \infty. \]
We use the parameterization $z=\E^{\I\phi}$,
$-\pi<\phi\le\pi$ of $S^1$. Recall that Hilbert
Schmidt operators $A\in{\cal J}_2(L^2(S^1))$ can be
written as square integrable kernels
$A(\phi,\phi')$. For instance, a rank-one-projection
$|e_r\rangle\langle e_r|$ has kernel $\E^{\I(\phi-\phi')}$.
For (small) $\epsilon>0$ define operators in
$S_{\rm NS}^{(\epsilon)},S_{\rm R}^{(\epsilon)}\in
{\cal J}_2(L^2(S^1))$ by kernels
\[ S_{\rm NS}^{(\epsilon)}(\phi,\phi') =
\sum_{n=0}^\infty
\E^{-(n+\frac{1}{2})(\I\phi-\I\phi'+\epsilon)}=
\frac{\E^{-\frac{1}{2}(\I\phi-\I\phi'+\epsilon)}}
{1-\E^{-(\I\phi-\I\phi'+\epsilon)}}, \]
and
\[ S_{\rm R}^{(\epsilon)}(\phi,\phi') =
\frac{1}{2} + \sum_{n=1}^\infty
\E^{-n(\I\phi-\I\phi'+\epsilon)}=
\frac{1}{1-\E^{-(\I\phi-\I\phi'+\epsilon)}}-
\frac{1}{2}, \]
Note that $\epsilon$ regularizes the singularities
for $\phi-\phi'=0,\pm 2\pi$. Using Cauchy's integral
formula, it is easy to check that for
$r,s \in \mathbb{Z} + \frac{1}{2}$
\begin{eqnarray*}
\lim_{\epsilon\searrow 0} \,\, \langle e_r,
S_{\rm NS}^{(\epsilon)} e_s \rangle &=&
\lim_{\epsilon\searrow 0} \oint_{S^1}
\frac{dz}{2\pi\I z} \oint_{S^1} \frac{dz'}{2\pi\I z'}\,\,
\frac{z^{-r-\frac{1}{2}} {z'}^{s+\frac{1}{2}}}
{1-\frac{z'}{z} \E^{-\epsilon}} \\
&=& \left\{ \begin{array}{cc}
\lim_{\epsilon\searrow 0} \,\, \E^{s\epsilon}\delta_{r,s}
\qquad & r,s<0 \\ 0 \qquad & \mbox{otherwise}
\end{array} \right.
= \langle e_r, S_{\rm NS} e_s \rangle .
\end{eqnarray*}
Because $\E^{s\epsilon}<1$ for $s<0$ this result
can be generalized to
\[ \lim_{\epsilon\searrow 0} \,\,
\langle f, S_{\rm NS}^{(\epsilon)} g \rangle =
\langle f, S_{\rm NS} g \rangle \]
for arbitrary $f,g \in L^2(S^1)$ by an argument
of bounded convergence. So we have weak convergence
$\mbox{w-}\lim_{\epsilon\searrow 0}
S_{\rm NS}^{(\epsilon)}=S_{\rm NS}$. In an
analogous way one obtains
$\mbox{w-}\lim_{\epsilon\searrow 0}
S_{\rm R}^{(\epsilon)}=S_{\rm R}$. Thus the
difference $\Delta^{(\epsilon)}=
S_{\rm R}^{(\epsilon)}-S_{\rm NS}^{(\epsilon)}$
with kernel
\[ \Delta^{(\epsilon)}(\phi,\phi')=
\frac{1}{1+\E^{-\frac{1}{2}(\I\phi-\I\phi'+\epsilon)}}-
\frac{1}{2} \]
converges weakly to $\Delta=S_{\rm R}-S_{\rm NS}$.
We have to show that $X=P_I\Delta P_I$ is trace class.
The operator $P_I$ acts as multiplication with
the characteristic function $\chi_I(\phi)$
corresponding to $z=\E^{\I\phi}\in I$. Now
$X^{(\epsilon)}=P_I \Delta^{(\epsilon)} P_I$,
converging weakly to $X$, has kernel
\[ X^{(\epsilon)}(\phi,\phi')= \chi_I(\phi) \left(
\frac{1}{1+\E^{-\frac{1}{2}(\I\phi-\I\phi'+\epsilon)}}-
\frac{1}{2} \right) \chi_I(\phi') \]
and is no more singular for $\epsilon\searrow 0$. Hence
\[ \lim_{\epsilon\searrow 0}\,\,
\langle f, X^{(\epsilon)} g \rangle
= \int_{-\pi}^\pi \frac{d\phi}{2\pi}
\int_{-\pi}^\pi \frac{d\phi'}{2\pi} \,\,
\overline{f(\E^{\I\phi})} X^{(\epsilon=0)}(\phi,\phi')
g(\E^{\I\phi'}), \qquad f,g\in L^2(S^1), \]
by the theorem of bounded convergence. It follows
$X=X^{(\epsilon=0)}\in{\cal J}_2(L^2(S^1))$.
Let $\tilde{\chi}_I$ be a smooth function on
$[-\pi,\pi]$ which satisfies $\tilde{\chi}_I(\phi)=1$
for $z=\E^{\I\phi}\in I$ and vanishes
in a neighborhood of $\phi=\pm\pi$. We define
\[ \tilde{X}(\phi,\phi')=\tilde{\chi}_I(\phi)
\left( \frac{1}{1+\E^{-\frac{\I}{2}(\phi-\phi')}}
-\frac{1}{2} \right) \tilde{\chi}_I(\phi') \]
such that $X=P_I\tilde{X}P_I$ and hence
\[ \|X\|_1=\|P_I\tilde{X}P_I\|_1 \le \|P_I \|
\|\tilde{X}\|_1 \|P_I\| = \|\tilde{X}\|_1. \]
Since $\tilde{X}(\phi,\phi')$ is a smooth function
in $\phi$ and $\phi'$ it has fast decreasing
Fourier coefficients which coincide with matrix
elements $\langle e_n,\tilde{X} e_m \rangle$,
$n,m\in\mathbb{Z}$. This proves the statement
$\|X\|_1\le\infty$, q.e.d.

We have proven local quasiequivalence of our
representations $\pi_0$, $\pi_{1/2}$ and $\pi_1$.
Thus we have an extension to local von Neumann
algebras ${\cal R}(I)$ and to the quasilocal algebra
$\mathfrak{A}_\zeta$ they generate. By unitary
equivalence $\varrho_J^{\rm loc}\cong\pi_J$ on
${\cal A}(I_\zeta)$ i.e.~there are unitaries
$U_J:{\cal H}_0\rightarrow{\cal H}_J$ satisfying
$\varrho_J^{\rm loc}(A)=U_J^{-1}\pi_J(A)U_J$ for all
$A\in{\cal A}(I)$ and all $I\in{\cal J}_\zeta$, we
have an extension of $\varrho_J^{\rm loc}$ to
$\mathfrak{A}_\zeta$, too, $J=\frac{1}{2},1$. These
extensions, denoted by the same symbols, fulfill
\[ \varrho_J^{\rm loc}(A)=A, \qquad
A\in{\cal R}(I_1),\quad I_1\in{\cal J}_\zeta,
\quad I_1\cap I=\emptyset \]
($I$ denotes the localization region) and also
\[ \varrho_J^{\rm loc}({\cal R}(I_0)) \subset
{\cal R}(I_0);\qquad I_0\in{\cal J}_\zeta,\quad
I\subset I_0, \]
because these endomorphisms satisfy the
corresponding relations on the underlying
$C^*$-algebras ${\cal A}(I_1)$ and ${\cal A}(I_0)$.
We have established that our endomorphisms
$\varrho_J^{\rm loc}$, $J=\frac{1}{2},1$, are
well-defined localized endomorphisms in the
common sense. In addition, these endomorphisms
are transportable. This follows because the
precosheaf $\{{\cal R}(I)\}$ is M{\"o}bius covariant.
Thus $\mathfrak{A}_\zeta$ is covariant with respect to
the subgroup of M{\"o}bius transformations leaving
$\zeta$, the point at infinity, invariant.

\subsection{Fusion Rules of Localized Endomorphisms}
The main advantage of working with local von Neumann
algebras is that one can manifestly prove fusion
rules in terms of equivalence classes of localized
endomorphisms. Let $\varrho_a$ and $\varrho_b$ be
endomorphisms of $\mathfrak{A}_\zeta$ localized in
intervals $I_a,I_b\in{\cal J}_\zeta$, respectively.
Then there is an interval $I\in{\cal J}_\zeta$,
$I_a\cup I_b\subset I$, such that $\varrho_a$ and
$\varrho_b$ are localized in $I$. Suppose that
$\varrho_a\cong\varrho_b$, i.e.~that there is a
unitary $U\in{\cal B}({\cal H}_0)$ such that
$\varrho_a(A)=U^*\varrho_b(A)U$ for all
$A\in\mathfrak{A}_\zeta$. Exploiting Haag duality one
finds $U\in{\cal R}(I)$, i.e. $U\in\mathfrak{A}_\zeta$.
This is an important tool which enables to derive
fusion rules in the algebraic framework:
If also $\tilde{\varrho}_a(A)=
\tilde{U}^*\tilde{\varrho}_b(A)\tilde{U}$,
$A\in\mathfrak{A}_\zeta$, for localized endomorphisms
$\tilde{\varrho}_a,\tilde{\varrho}_b$ then
$\tilde{\varrho}_a\varrho_a\cong\tilde{\varrho}_b
\varrho_b$, realized by the well-defined unitary
$\tilde{\varrho}_b(U)\tilde{U}$. So we can deduce
fusion rules in terms of equivalence classes by
computing it for some special representatives.
Obviously, this procedure fails for global
endomorphisms $\varrho_J$, $J=\frac{1}{2},1$,
of ${\cal A}_{\rm univ}^{C^*}$ by two reasons:
The first one is that ${\cal A}_{\rm univ}^{C^*}$
is generated by local $C^*$-algebras. But local
intertwiners may lie only in their weak closures;
there is a rather small number of endomorphisms
which are inner equivalent in the $C^*$-algebras.
The second reason is that the vacuum representation
does not act faithfully on ${\cal A}_{\rm univ}^{C^*}$.
In the vacuum representation, too much information gets
lost. For example, the equivalence class of
$\pi_0\circ\varrho_{1/2}$ does not depend on the
isometry $V_{1/2}'$ at all but the representation
$\pi_0\circ\varrho_{1/2}^2$ does.
Counterexamples can be constructed; there is, for
instance, an endomorphism $\mu_{1/2}$ such that
$\pi_0\circ\mu_{1/2}\cong\pi_0\circ\varrho_{1/2}$,
but $\pi_0\circ\mu_{1/2}^2\not\cong
\pi_0\circ\varrho_{1/2}^2$ as representations
of ${\cal A}_{\rm univ}^{C^*}$ \cite{ich}.

It is no problem to compute the fusion rules
for our special examples of localized
endomorphisms, we just have to summarize
some of our previous results. Proposition
\ref{square} gives us the first fusion rule,
we have established
\[ \pi_0\circ\varrho_{1/2}^{\rm loc}
\varrho_{1/2}^{\rm loc}\cong\pi_0\oplus\pi_1. \]
We obtain the second fusion rule by the fact that
$\pi_1\circ\varrho_{1/2}^{\rm loc}\cong
\pi_{1/2}$ (Theorem \ref{1/2loc}). Hence we conclude
\[ \pi_0\circ\varrho_1^{\rm loc}
\varrho_{1/2}^{\rm loc} \cong
\pi_1\circ\varrho_{1/2}^{\rm loc}\cong
\pi_{1/2}.\]
Since $\varrho_1^{\rm loc}$ and
$\varrho_{1/2}^{\rm loc}$ commute if we choose
the localization region of $\varrho_1^{\rm loc}$
disjoint to that of $\varrho_{1/2}^{\rm loc}$
we also obtain
\[ \pi_0\circ\varrho_{1/2}^{\rm loc}
\varrho_1^{\rm loc}\cong\pi_{1/2}.\]
Trivially, the fact that $(\varrho_1^{\rm loc})^2=id$ leads
us to the third fusion rule
\[ \pi_0 \circ \varrho_1^{\rm loc} \varrho_1^{\rm loc} \cong
\pi_0. \]
Denoting by $\varrho_0$ the identity endomorphism
(everywhere localized) and by $[\varrho_J]$ the
equivalence class of localized endomorphisms being
unitarily equivalent to $\varrho_J^{\rm loc}$ in the
vacuum representation, we summarize
\begin{theorem}[Fusion rules of localized endomorphisms]
\begin{eqnarray}
{[} \varrho_{1/2}^2 {]} &=&
{[} \varrho_0 {]} + {[} \varrho_1 {]} ,\\
{[} \varrho_{1/2} \varrho_1 {]} =
{[} \varrho_1 \varrho_{1/2} {]}
&=& {[} \varrho_{1/2} {]} ,\\
{[} \varrho_1^2 {]} &=& {[} \varrho_0 {]},
\end{eqnarray}
\end{theorem}
i.e. the localized endomorphisms obey the Ising
fusion rules.

\subsection{Statistics Operator and Left Inverse}
According to the general theory of superselection
sectors \cite{DHR2,FRS1,Haag}, we expect that for each
endomorphism $\varrho$ which is localized in some
interval $I\in{\cal J}_\zeta$ there exists a
unitary $\varepsilon_\varrho\in{\cal R}(I)$
which commutes with $\varrho^2(\mathfrak{A}_\zeta)$,
\[ \varepsilon_\varrho\in\varrho^2(\mathfrak{A}_\zeta)'\]
and fulfills
\begin{equation}
\label{braid}
\varepsilon_\varrho \varrho (\varepsilon_\varrho)
\varepsilon_\varrho = \varrho (\varepsilon_\varrho)
\varepsilon_\varrho \varrho (\varepsilon_\varrho).
\end{equation}
Therefore the elements $\tau_i=\varrho^{i-1}
(\varepsilon_\varrho)$, $i=1,2,\ldots,$ satisfy the
Artin relations and determine a representation of
the braid group $B_\infty$ \cite{DHR2,FRS1}.
The statistics operator is given by the formula
\begin{equation}
\label{stat}
\varepsilon_\varrho = U^{-1}\varrho(U)
\end{equation}
where $U$ is unitary such that the (equivalent)
endomorphism $\tilde{\varrho}$, defined by
\[ \tilde{\varrho}(A)=U\varrho(A)U^{-1},\qquad
A\in\mathfrak{A}_\zeta \]
is localized in some interval
$I_0\in{\cal J}_\zeta$, $I_0\subset I'$. The
statistics operator is independent of the special
choice of $\tilde{\varrho}$ as far as $I_0$ varies
in one of the two connected components of
$I'\setminus\{\zeta\}$ but it may depend on
the fact whether $I_0$ lies in the left or the
right complement of $I$ with respect to our ``point
at infinity'' $\zeta$. The computation of
$\varepsilon_\varrho$ is straightforward for
$\varrho=\varrho_1^{\rm loc}$. Let $\varrho$ be
induced by a real function $h\in L^2(S^1)$ with
support in some interval $I$, $\|h\|^2=1$,
as described in Definition \ref{rho1loc}.
Analogously, let $\tilde{\varrho}$
be induced by a real function $h_0$, $\|h_0\|=1$,
with supp$(h_0)\subset I_0$, $I_0\cap I=\emptyset$.
Since these endomorphisms
are unitarily implemented we find
\[ U=2\psi(h_0)\psi(h),\qquad U^{-1}=
2\psi(h)\psi(h_0) \]
and
\[ \varrho(U) = 2\psi(-h_0)\psi(h)= -U \]
so that
\[ \varepsilon_\varrho = - {\bf 1} \]
expressing nothing but anticommutativity of
Majorana fields.
We now want to construct the statistics operator
$\varepsilon_\sigma$ for our localized endomorphism
$\sigma=\sigma_{1/2}^{\rm loc}$. It seems to be
very difficult to do that by the formula (\ref{stat})
but to be much easier to determine it by its properties.
The statistics operator commutes with
$\sigma^2(\mathfrak{A}_\zeta)$. The commutant
$\sigma^2(\mathfrak{A}_\zeta)'$ is spanned by
elements ${\bf 1},\Pi$ where the projection $\Pi$
is defined by
\[ \Pi = \psi(e_+)\psi(e_-),  \quad e_\pm=
\frac{1}{\sqrt{2}}\left( e_0^{(2)} \pm \I
V'e_0^{(2)} \right). \]
(We remark that the orthonormal vectors $e_\pm$ span
the kernel of $(V'V')^*$ and satisfy $e_+=\Gamma e_-$.)
This leads us to the ansatz
\[ \varepsilon_\sigma = \alpha ({\bf 1}+\gamma \Pi),
\qquad \alpha,\gamma \in \mathbb{C}. \]
Now $\varepsilon_\sigma$ is unitary,
\[ \varepsilon_\sigma \varepsilon_\sigma^* =
|\alpha|^2({\bf 1} + (\gamma + \bar{\gamma} +
\gamma \bar{\gamma} ) \Pi ) = {\bf 1}. \]
Therefore $\gamma + \bar{\gamma} +
\gamma \bar{\gamma}=0$, $|\alpha|^2=1$, we write
$\alpha=\E^{\I\omega}$, $\omega$ real. The statistics
operator satisfies Eq.~(\ref{braid});
we exclude the case $\gamma=0$ and find
\begin{eqnarray*}
0 &=& \E^{-3\I \omega} \gamma^{-1} (\varepsilon_\sigma
\sigma(\varepsilon_\sigma) \varepsilon_\sigma -
\sigma(\varepsilon_\sigma) \varepsilon_\sigma
\sigma(\varepsilon_\sigma)) \\
&=& (\gamma+1)(\Pi-\sigma(\Pi)) + \gamma^2(\Pi
\sigma(\Pi)\Pi - \sigma(\Pi)\Pi\sigma(\Pi)).
\end{eqnarray*}
It is not hard to see that $\Pi$ and $\sigma(\Pi)$
can be written in the following way:
\[ \Pi=\frac{1}{2}\left( {\bf 1} + 2\I \psi
(V'e_0^{(2)})\psi(e_0^{(2)}) \right), \quad
\sigma(\Pi)= \frac{1}{2}\left( {\bf 1}
+ 2\I \psi((V')^2 e_0^{(2)})\psi(V'e_0^{(2)})\right). \]
The fields obey
\[ \psi(e_0^{(2)})^2 = \psi(V'e_0^{(2)})^2 =
\psi((V')^2 e_0^{(2)})^2 = \frac{1}{2} {\bf 1}, \]
and
\[ \{ \psi(V'e_0^{(2)}), \psi(e_0^{(2)}) \} =
\{ \psi((V')^2 e_0^{(2)}),\psi(e_0^{(2)}) \} =
\{ \psi((V')^2 e_0^{(2)}),\psi(V' e_0^{(2)})\}=0.\]
Using these relations one finds\footnote{With
the identification $E_n=\sigma^{n-1}(\Pi)$,
$n=1,2,\ldots$, this is nothing else but the
Temperley-Lieb-Jones algebra relation
\[ E_n E_{n\pm1} E_n = d(\sigma)^{-2} E_n \]
with statistical dimension $d(\sigma)=\sqrt{2}$.}
\[ \Pi \sigma(\Pi) \Pi = \frac{1}{2} \Pi, \qquad
\sigma(\Pi)\Pi\sigma(\Pi)=\frac{1}{2}\sigma(\Pi)\]
so that we obtain
\[ \left( \gamma + 1 + \frac{1}{2} \gamma^2 \right)
\left( \Pi - \sigma(\Pi) \right) = 0. \]
Since $\Pi-\sigma(\Pi)\neq 0$ we have
\[ \gamma^2 + 2\gamma + 2 =0 \qquad
\Longleftrightarrow \qquad \gamma = -1 \pm \I \]
and therefore
\[ \varepsilon_\sigma = \E^{\I\omega} ({\bf 1}
-(1\pm i)\Pi). \]
According to the general theory
\cite{DHR2,FRS1,Haag}, we expect that there
exists also a left inverse $\Phi_\sigma$ to our
endomorphism $\sigma$ such that $\Phi_\sigma\circ
\sigma=id$. The left inverse is a unital, positive
mapping from $\mathfrak{A}_\zeta$ to
$\mathfrak{A}_\zeta$ which satisfies
$\Phi_\sigma({\cal R}(I))\subset {\cal R}(I)$ if
$I\supset I_2$. Since $\sigma$ is not an automorphism
$\Phi_\sigma$ does in general not respect products but
\[ \Phi_\sigma(\sigma(A)B\sigma(C))=A\Phi_\sigma(B)C \]
holds for $A,B,C\in\mathfrak{A}_\zeta$. In the following
we want to derive an explicit description for
$\Phi_\sigma$. We introduce an arbitrary orthonormal
basis $\{ v_n,n\in\mathbb{Z}\}$ of $L^2(S^1)$ with
$v_0=e_0^{(2)}$ and $\Gamma v_n = v_{-n}$.
It suffices to consider elements $A$ of $\mathfrak{A}_\zeta$
which are sums of monomials $X$ of the form
\[ X=\psi(v_{n_1}) \psi(v_{n_2}) \cdots
\psi(v_{n_{2K}}). \]
Using the anticommutation relations,
\[ \{ \psi(v_n),\psi(v_{n'})\}=\delta_{n,-n'}{\bf 1},\]
in particular $\psi(v_0)^2=\frac{1}{2}{\bf 1}$,
we can write every monomial $X$ such that $\psi(v_0)$
appears at most once. If every monomial $X$ is written
in that way we define $\Phi_\sigma$ as the linear mapping
which preserves the unit and fulfills
\[ \Phi_\sigma(X) = \psi({V'}^*v_{n_1})\psi({V'}^*v_{n_2})
\cdots \psi({V'}^*v_{n_{2K}}). \]
It is no problem to check that $\Phi_\sigma$ is well
defined and has the required properties. The general
theory says
\[ \Phi_\sigma (\varepsilon_\sigma) = \frac{\omega_\sigma}
{d(\sigma)} {\bf 1} \]
where $\omega_\sigma$ is a phase factor (``statistical
phase'') and the positive real number $d(\sigma)$ is
called statistical dimension. Since
$\sigma=\sigma_{1/2}^{\rm loc}$ belongs to the
sector $[\varrho_{1/2}]$ we expect that
$d(\sigma)=\sqrt{2}$. Using our formula
for $\Phi_\sigma$ we find (respecting that
$V'e_0^{(2)}$ is orthogonal to $e_0^{(2)}$)
\[ \Phi_\sigma(\Pi) = \Phi_\sigma \left(
\frac{1}{2} \left({\bf 1} + 2\I \psi
(V'e_0^{(2)})\psi(e_0^{(2)})
\right)\right) = \frac{1}{2}{\bf 1}. \]
We conclude
\[ \Phi_\sigma(\varepsilon_\sigma)= \E^{\I\omega}
\left( 1- \left( \frac{1}{2} \pm \frac{\I}{2}
\right)\right) {\bf 1} = \frac{\E^{\I\left(\omega
\mp \frac{\pi}{4} \right)}}{\sqrt{2}} {\bf 1}, \]
in agreement with $d(\sigma)=\sqrt{2}$. At the end
we find
\begin{equation}
\varepsilon_\sigma= \frac{\omega_\sigma}{\sqrt{2}}
((1\pm \I){\bf 1} \mp 2\I \Pi).
\end{equation}
By the spin and statistics theorem \cite{FRS2} we
expect that the statistical phase is given by
$\omega_\sigma=\E^{2\pi\I s}$ where $s$ is the infimum
of the spectrum of the conformal energy operator $L_0$
in the representation $\pi_0\circ\sigma$. Since
$\sigma$ belongs to the sector $[\varrho_{1/2}]$ we
have $s=\frac{1}{16}$ and therefore
$\omega_\sigma=\E^\frac{\I\pi}{8}$. However,
we did not succeed in computing $\omega_\sigma$
directly. Moreover, we observe the freedom to choose
the $\pm$-sign in our formula for the statistics
operator $\varepsilon_\sigma$. The change of this
sign corresponds to the replacement of
$\varepsilon_\sigma$ by $\varepsilon_\sigma^*$.
The fact that
$\varepsilon_\sigma\neq\varepsilon_\sigma^*$
goes back to the non-trivial spacetime topology
which is the origin of braid statistics. At the
end we remark that the same calculations we have
done for $\sigma=\sigma_{1/2}^{\rm loc}$ run for the
endomorphism $\varrho_{1/2}^{\rm loc}$; we just have
to replace $V'$ by $V$ and $e_0^{(2)}$ by $f_0^{(2)}$.

\begin{appendix}
\section{Appendix: M{\"o}bius Covariance of the
Vacuum Sector}
We will briefly discuss M{\"o}bius covariance
here. Related topics can be found in the book
of Lang \cite{Lang}. The M{\"o}bius symmetry on
the circle $S^1$ is given by the group
${\rm Mob}={\rm SU}(1,1)/\mathbb{Z}_2$ where
\[ {\rm SU}(1,1) = \left\{ \left. g= \left(
\begin{array}{cc}
\alpha & \beta \\ \overline{\beta} & \overline{\alpha}
\end{array} \right) \in {\rm GL}_2(\mathbb{C}) \,\,
\right| \,\, |\alpha|^2-|\beta|^2=1 \right\}. \]
Its action on the circle is
\[ gz = \frac{\overline{\alpha}z-\overline{\beta}}{-\beta
z+\alpha}, \qquad z\in S^1. \]
Each element $g\in{\rm SU}(1,1)$ can be decomposed
in the product of a rotation $r(t)$ and a transformation
$g'=r(t)^{-1}g$ leaving the point $z=-1$ invariant,
\[ g=r(t)g',\qquad r(t)= \left( \begin{array}{cc}
\E^{-\frac{\I t}{2}} & 0 \\ 0 & \E^\frac{\I t}{2}
\end{array} \right), \qquad t\in\mathbb{R},
\qquad g'=\left( \begin{array}{cc}
\alpha' & \beta' \\ \overline{\beta'} &
\overline{\alpha'} \end{array} \right) , \]
such that $(\overline{\alpha'}+\overline{\beta'})
(\alpha'+\beta')^{-1}=1$.
Since $r(t+2\pi)=-r(t)$ we can determine
$-2\pi<t\le 2\pi$ uniquely by the additional
requirement ${\rm Re}(\alpha')>0$. Then a
representation $U$ of ${\rm SU}(1,1)$
in our Hilbert space of test functions $L^2(S^1)$ is
defined by
\begin{equation}
\label{repr}
\left( U (g) f \right) (z) = \epsilon(g;z)
(\alpha + \overline{\beta}
\overline{z})^{-\frac{1}{2}}(\overline{\alpha}
+\beta z)^{-\frac{1}{2}}
f\left( \frac{\alpha z + \overline{\beta}}{\beta z
+ \overline{\alpha}} \right)
\end{equation}
where for $z=\E^{\I\phi}$, $-\pi<\phi\le\pi$,
\[ \epsilon(g;z) = - \, {\rm sign} (t-\pi-\phi)
\,\, {\rm sign}(t+\pi-\phi), \]
and ${\rm sign(x)}=1$ if $x\ge 0$,
${\rm sign}(x)=-1$ if $x<0$. We observe that
$\epsilon(g;z)$ is discontinuous at $z=-1$
and $z=g(-1)=-(\overline{\alpha}+\overline{\beta})
(\alpha+\beta)^{-1}$.
Up to this $\epsilon$-factor, Eq.~(\ref{repr})
is a well-known definition of a representation
of ${\rm SU}(1,1)$. So it remains to be checked
that
\[ \epsilon(g_1;z) \epsilon(g_2;g_1^{-1}z)
=\epsilon(g_1g_2;z). \]
Since both sides have their discontinuities at
$z=-1$ and $z=g_1g_2(-1)$ they can differ only
by a global sign. But this possibility is easily
excluded by arguments of $L^2$-continuity in
$g$. Moreover,
by computing $\langle U(g) e_r, U(g) e_s \rangle
= \delta_{r,s}$ for $r,s\in\mathbb{Z}+\frac{1}{2}$
(NS-base) we can also check that $U$ is unitary,
\begin{eqnarray*}
\langle U(g) e_r , U(g) e_s \rangle &=&
\oint_{S^1} \frac{dz}{2\pi\I z}
(\alpha + \overline{\beta}\overline{z})^{-1}
(\overline{\alpha} + \beta z)^{-1}
\left( \frac{\alpha z + \overline{\beta}}
{\beta z + \overline{\alpha}} \right)^{s-r} \\
&=& \frac{1}{2\pi\I} \oint_{S^1} dz \,\,
(\alpha z + \overline{\beta})^{s-r-1}
(\beta z + \overline{\alpha})^{r-s-1} \\
&=& \left\{ \begin{array}{cc}
0 & \quad s>r \\
\frac{\alpha^{s-r-1}}{(r-s)!}
\frac{d^{r-s}}{dz^{r-s}}(\beta z+
\overline{\alpha})^{r-s-1}\big|_{z=-
\frac{\overline{\beta}}{\alpha}} &
\quad s \le r \end{array} \right. \\
&=& \delta_{r,s}
\end{eqnarray*}
by Cauchy's integral formula, respecting that
$|\alpha|^2>|\beta|^2$ since $|\alpha|^2-
|\beta|^2=1$. Since the prefactor
on the right hand side in Eq.~(\ref{repr})
is real we observe $[U(g),\Gamma]=0$ and hence
each $U(g)$, $g\in{\rm SU}(1,1)$ induces a
Bogoliubov automorphism $\alpha_g=\varrho_{U(g)}$
of ${\cal C}(L^2(S^1),\Gamma)$. Hence
${\rm SU}(1,1)$ is represented by automorphisms
of ${\cal C}(L^2(S^1),\Gamma)$, and this restricts
to a representation of ${\rm Mob}$ by automorphisms
of ${\cal C}(L^2(S^1),\Gamma)^+$.
In order to establish
M{\"o}bius invariance of the vacuum state and hence
covariance of the vacuum sector we show that
\[ [S_{\rm NS},U(g)]=0,\qquad g\in{\rm SU}(1,1),\]
i.e.~that $U(g)$ respects the polarization of
$L^2(S^1)$ induced by $S_{\rm NS}$. It is
sufficient to show that
\[ \langle e_{-r},U(g) e_s \rangle =0,
\quad r,s\in\mathbb{N}_0 + \frac{1}{2}, \qquad
g \in {\rm SU}(1,1). \]
The functions $e_r(z)$, $r\in\mathbb{Z}+\frac{1}{2}$
are smooth on $S^1$ except at their cut at
$z=-1$. The prefactor $\epsilon(g;z)$ in
Eq.~(\ref{repr}) achieves that
$(U(g)e_r)(z)$ remains a smooth function except
at $z=-1$, i.e. that the cut is not transported
to $g(-1)$. Hence we have
\[ (U(g)e_r)(z)= \pm(\alpha
+\overline{\beta}\overline{z})^{-\frac{1}{2}}
(\overline{\alpha}+\beta z
)^{-\frac{1}{2}} \left( \frac{\alpha z +
\overline{\beta}}{\beta z + \overline{\alpha}}
\right)^r \]
where all the half-odd integer powers are to be
taken in the same branch with cut at
$z=-1$. So we can compute as follows,
\begin{eqnarray*}
\langle e_{-r}, U(g) e_s \rangle &=& \pm
\oint_{S^1} \frac{dz}{2\pi\I z} z^r
(\alpha z+\overline{\beta})^{-\frac{1}{2}}
z^\frac{1}{2}(\overline{\alpha}+\beta z
)^{-\frac{1}{2}} \left( \frac{\alpha z +
\overline{\beta}}{\beta z + \overline{\alpha}}
\right)^s \\
&=& \pm\frac{1}{2\pi\I} \oint_{S^1} dz \,\,
z^{r-\frac{1}{2}} (\alpha z +\overline{\beta}
)^{s-\frac{1}{2}} (\overline{\alpha} +
\beta z)^{-s-\frac{1}{2}} = 0,
\end{eqnarray*}
again by Cauchy's formula, respecting
$|\alpha|^2>|\beta|^2$ and that $r,s$ are
positive half-odd integers here.

\section{Appendix: The Proof of Lemma 5.3}
An essential fact we use for the proof of Lemma
\ref{HS} is presented in the following
\begin{lemma}
The difference of the two odd pseudolocalized Bogoliubov
operators, given in Definition \ref{def1/2loc} is
Hilbert Schmidt class,
\begin{equation}
V-V' \in {\cal J}_2 (L^2(S^1)).
\end{equation}
\end{lemma}
{\it Proof.} Since ${\bf 1}-P_0^{(2)}$ is Hilbert Schmidt
class, where
\[ P_0^{(2)}= {\bf 1} - |e_0^{(2)} \rangle\langle e_0^{(2)}|
= P_{I_+}+P_{I_-} + \sum_{n=1}^\infty \Big( |e_n^{(2)}
\rangle\langle e_n^{(2)} | + |e_{-n}^{(2)} \rangle\langle
e_{-n}^{(2)}| \Big), \]
it is equivalent to prove
\[ \Sigma_0 = \| P_0^{(2)}(V-V')P_0^{(2)}
\|_2^2 < \infty. \]
We remember that the square of the Hilbert Schmidt norm
is the sum over the squares of all matrix elements in any
Hilbert space basis. Obviously, the Bogoliubov operators
$V$ and $V'$ differ only on the subspace
$L^2(I_2)\subset L^2(S^1)$. We compute
\begin{eqnarray*}
\Sigma_0 &=& \sum_{n\in\mathbb{Z}\atop n\neq 0}
\sum_{m\in\mathbb{Z}\atop m\neq 0} \Big|
\langle e_n^{(2)}, V e_m^{(2)} \rangle -
\langle e_n^{(2)}, V'e_m^{(2)} \rangle \Big|^2 \\
&=& \sum_{n=1}^\infty \sum_{m=1}^\infty \Big|
\I \langle e_n^{(2)}, e_{m+\frac{1}{2}}^{(2)} \rangle -
\I \langle e_{n-\frac{1}{2}}^{(2)},e_m^{(2)} \rangle
\Big|^2 \\
&& \qquad + \sum_{n=1}^\infty \sum_{m=-1}^{-\infty} \Big|
- \I \langle e_n^{(2)}, e_{m-\frac{1}{2}}^{(2)} \rangle
- \I \langle e_{n-\frac{1}{2}}^{(2)},e_m^{(2)} \rangle
\Big|^2 \\
&& \qquad + \sum_{n=-1}^{-\infty} \sum_{m=1}^\infty \Big|
\I \langle e_n^{(2)}, e_{m+\frac{1}{2}}^{(2)} \rangle
+ \I \langle e_{n+\frac{1}{2}}^{(2)},e_m^{(2)} \rangle
\Big|^2 \\
&& \qquad + \sum_{n=-1}^{-\infty} \sum_{m=-1}^{-\infty} \Big|
- \I \langle e_n^{(2)}, e_{m-\frac{1}{2}}^{(2)} \rangle
+ \I \langle e_{n+\frac{1}{2}}^{(2)},e_m^{(2)} \rangle
\Big|^2.
\end{eqnarray*}
Since $\langle e_{n\pm\frac{1}{2}}^{(2)}, e_m^{(2)}
\rangle = \langle e_n^{(2)}, e_{m\mp\frac{1}{2}}^{(2)}
\rangle$ the first and the fourth summation vanishes,
so that one finds by substituting to positive
summation indices
\begin{eqnarray*}
\Sigma_0 &=& \sum_{n=1}^\infty \sum_{m=1}^\infty \Big|
\langle e_n^{(2)}, e_{-m-\frac{1}{2}}^{(2)} \rangle
+ \langle e_{n-\frac{1}{2}}^{(2)},e_{-m}^{(2)} \rangle
\Big|^2 \\ && \qquad
+ \sum_{n=1}^\infty \sum_{m=1}^\infty \Big|
\langle e_{-n}^{(2)}, e_{m+\frac{1}{2}}^{(2)} \rangle
+ \langle e_{-n+\frac{1}{2}}^{(2)},e_m^{(2)} \rangle
\Big|^2 \\
&=& 2 \sum_{n=1}^\infty \sum_{m=1}^\infty \Big|
\langle e_{-n}^{(2)}, e_{m+\frac{1}{2}}^{(2)} \rangle
+ \langle e_{-n+\frac{1}{2}}^{(2)},e_m^{(2)} \rangle
\Big|^2,
\end{eqnarray*}
we used $\langle e_{-n+\frac{1}{2}}^{(2)},e_m^{(2)} \rangle
= \overline{\langle e_{n-\frac{1}{2}}^{(2)}, e_{-m}^{(2)}
\rangle}$. The remaining matrix elements are easily
computed,
\begin{eqnarray*}
\langle e_{-n}^{(2)}, e_{m+\frac{1}{2}}^{(2)} \rangle &=&
2 \int_{-\frac{\pi}{2}}^\frac{\pi}{2} \E^{2\I (m+n+\frac{1}{2})
\phi} \frac{d\phi}{2\pi} = + \frac{(-1)^{m+n}}{\pi} \,
\frac{1}{m+n+\frac{1}{2}} , \\
\langle e_{-n+\frac{1}{2}}^{(2)}, e_m^{(2)} \rangle &=&
2 \int_{-\frac{\pi}{2}}^\frac{\pi}{2} \E^{2\I (m+n-\frac{1}{2})
\phi} \frac{d\phi}{2\pi} = - \frac{(-1)^{m+n}}{\pi} \,
\frac{1}{m+n-\frac{1}{2}}.
\end{eqnarray*}
It follows
\[ \Sigma_0 = \frac{2}{\pi^2} \sum_{n=1}^\infty
\sum_{m=1}^\infty \frac{1}{\left( (n+m)^2 - \frac{1}{4}
\right)^2} = \frac{2}{\pi^2} \sum_{k=1}^\infty
\frac{k}{\left( (k+1)^2 - \frac{1}{4} \right)^2}
< \infty, \qquad \mbox{q.e.d.} \]
Now we can start proving Lemma \ref{HS}.
We introduce the following notations
\begin{eqnarray*}
B &=& V^*S_{\rm NS}V-S_{\rm R},\\
P_0 &=& |e_{-1}\rangle\langle e_{-1}| +
|e_0 \rangle\langle e_0 | + |e_1 \rangle\langle e_1|,\\
P_1 &=&\sum_{n=1}^\infty |e_{-2n-1}\rangle\langle e_{-2n-1}|,\\
P_2 &=&\sum_{n=1}^\infty |e_{2n+1}\rangle\langle e_{2n+1}|,\\
P_3 &=&\sum_{n=1}^\infty |e_{-2n}\rangle\langle e_{-2n}|,\\
P_4 &=&\sum_{n=1}^\infty |e_{2n}\rangle\langle e_{2n}|,
\end{eqnarray*}
such that we find
\[ \sum_{i=0}^4 P_i = {\bf 1}, \qquad
\Gamma P_1 = P_2 \Gamma, \qquad
\Gamma P_3 = P_4 \Gamma. \]
At first we have to show, that $\|B\|_2<\infty$. Since
$P_0$ is Hilbert Schmidt class it is equivalent to
prove that
\[ \| ({\bf 1}-P_0)B({\bf 1}-P_0) \|_2 =
\left\| \sum_{i,j=1}^4 P_i B P_j \right\|_2 \le
\sum_{i,j=1}^4 \|P_i B P_j \|_2 < \infty. \]
This will be done by estimating each term $\|P_iBP_j\|_2$
for its own. Since $B=B^*$ we find
\[ \| P_i B P_j \|_2 =\|(P_i B P_j)^* \|_2 =
\|P_j B P_i\|_2,\]
so that we are allowed to treat only those ten of
sixteen terms with $i\le j$. Further, by
\[ \Gamma B \Gamma = V^* \Gamma S_{\rm NS} \Gamma V - \Gamma S_{\rm R}
\Gamma =V^* ({\bf 1} - S_{\rm NS}) V - ({\bf 1}-S_{\rm R}) = -B \]
we find the identity
\[ \|P_1BP_1\|_2 = \| \Gamma P_1BP_1 \Gamma \|_2 =
\|P_2 \Gamma B \Gamma P_2 \|_2 = \|P_2BP_2\|_2, \]
and in the same way
\[ \| P_3 B P_3 \|_2  = \| P_4 B P_4 \|_2,\quad
\| P_2 B P_3 \|_2 = \| P_1 B P_4 \|_2, \quad
\| P_1 B P_3 \|_2 = \| P_2 B P_4 \|_2. \]
In each term on the right hand side one of the projections
$P_2$ or $P_4$ appears, but since
\[ P_2 S_{\rm R} = S_{\rm R} P_2 = P_4 S_{\rm R} = S_{\rm R} P_4 = 0 \]
we have only to prove the finiteness of the six norms
\begin{eqnarray*}
\|P_2 V^* S_{\rm NS} V P_2 \|_2,\quad
\|P_1 V^* S_{\rm NS} V P_2 \|_2,\quad
\|P_1 V^* S_{\rm NS} V P_4 \|_2,\\
\|P_2 V^* S_{\rm NS} V P_4 \|_2,\quad
\|P_3 V^* S_{\rm NS} V P_4 \|_2,\quad
\|P_4 V^* S_{\rm NS} V P_4 \|_2,
\end{eqnarray*}
and, since $V-V'$ is Hilbert Schmidt class, this is
equivalent to prove the finiteness of
\begin{eqnarray*}
\|P_2 {V'}^* S_{\rm NS} V' P_2 \|_2,\quad
\|P_1 {V'}^* S_{\rm NS} V' P_2 \|_2,\quad
\|P_1 {V'}^* S_{\rm NS} V P_4 \|_2,\\
\|P_2 {V'}^* S_{\rm NS} V P_4 \|_2,\quad
\|P_3 V^* S_{\rm NS} V P_4 \|_2,\quad
\|P_4 V^* S_{\rm NS} V P_4 \|_2.
\end{eqnarray*}
At first we consider
\begin{eqnarray*} \Sigma_1 &=& \|P_2{V'}^*S_{\rm NS}V'P_2\|_2^2 =
\sum_{n=1}^\infty \sum_{m=1}^\infty | \langle e_{2n+1},
{V'}^* S_{\rm NS} V' e_{2m+1} \rangle |^2 \\
&=& \sum_{n=1}^\infty \sum_{m=1}^\infty \left|
\sum_{r \in \mathbb{N}_0 + \frac{1}{2} }
\overline{\langle e_{-r},V' e_{2n+1} \rangle} \langle
e_{-r}, V' e_{2m+1} \rangle \right|^2.
\end{eqnarray*}
Since $\langle e_{m+\frac{1}{2}}^{(2)}, e_{2n+1}\rangle =
2^{-\frac{1}{2}} \delta_{n,m}$ the action of $V'$ on odd
basis vectors $e_{2n+1}$ is simple, one reads by
definition
\[ (V'e_{2n+1})(z) = \left\{ \begin{array}{cc} e_{2n+1}(z) &
\qquad z \in I_- \\
\I e_{2n+2}(z) & \qquad z \in I_2 \\ -e_{2n+1}(z) &
\qquad z \in I_+
\end{array} \right. \qquad n\in\mathbb{N}. \]
This leads us to
\begin{eqnarray*}
\lefteqn{\langle e_{-r},V' e_{2n+1} \rangle =}\\
&=& \int_{-\pi}^{-\frac{\pi}{2}}
\E^{\I(2n+1+r)\phi}\frac{d\phi}{2\pi} +
\I \int_{-\frac{\pi}{2}}^\frac{\pi}{2}
\E^{\I(2n+2+r)\phi} \frac{d\phi}{2\pi} -
\int_\frac{\pi}{2}^\pi \E^{\I(2n+1+r)\phi}
\frac{d\phi}{2\pi} \\
&=&  \frac{\I (-1)^n}{\pi} \sin
\left( \frac{r \pi}{2} \right) \frac{1}{(2n+1+r)(2n+2+r)}.
\end{eqnarray*}
Substituting to integer summation indices we obtain
\[ \Sigma_1 = \frac{64}{\pi^4} \sum_{n=1}^\infty
\sum_{m=1}^\infty \left( \sum_{l=0}^\infty
\sigma_{n,m,l}^{(1)} \right)^2 \]
where
\[ \sigma_{n,m,l}^{(1)} = \frac{1}{(4n+2l+3)(4n+2l+5)
(4m+2l+3)(4m+2l+5)}. \]
We put off the estimate of this summation for some
time and pass over to the next sum.
\[ \Sigma_2 = \| P_1 {V'}^* S_{\rm NS} V' P_2 \|_2^2
= \sum_{n=1}^\infty \sum_{m=1}^\infty \left|
\sum_{r \in \mathbb{N}_0 +\frac{1}{2}}
\overline{ \langle e_{-r}, V' e_{-2n-1} \rangle }
\langle e_{-r}, V' e_{2m+1} \rangle \right|^2 .\]
The action of $V'$ on vectors $e_{-2n-1}$ is
\[ (V'e_{-2n-1})(z) = \left\{ \begin{array}{cc}
e_{-2n-1}(z) & \qquad z \in I_- \\
-\I e_{-2n-2}(z) & \qquad z \in I_2 \\ -e_{-2n-1}(z) &
\qquad z \in I_+
\end{array} \right. \qquad n\in\mathbb{N} .\]
This leads us to
\[ \langle e_{-r}, V e_{-2n-1} \rangle = \frac{\I (-1)^n}{\pi}
\sin \left( \frac{r \pi}{2} \right)
\frac{1}{(2n+1-r)(2n+2-r)}, \]
so that
\[ \Sigma_2 = \frac{64}{\pi^4} \sum_{n=1}^\infty
\sum_{m=1}^\infty  \left( \sum_{l=0}^\infty
\sigma_{n,m,l}^{(2)} \right)^2  \]
where
\[ \sigma_{n,m,l}^{(2)} = \frac{1}{(4n-2l+1)(4n-2l+3)
(4m+2l+3)(4m+2l+5)}. \]
Analogously,
\[ \Sigma_3 = \| P_1 {V'}^* S_{\rm NS} V P_4 \|_2^2
= \sum_{n=1}^\infty \sum_{m=1}^\infty \left|
\sum_{r \in \mathbb{N}_0 +\frac{1}{2}}
\overline{ \langle e_{-r}, V' e_{-2n-1} \rangle }
\langle e_{-r}, V e_{2m} \rangle \right|^2 .\]
The action of $V$ on basis vectors $e_{2m}$ is
\[ (Ve_{2n})(z) = \left\{ \begin{array}{cc} e_{2n}(z) &
\qquad z \in I_- \\
\I e_{2n+1}(z) & \qquad z \in I_2 \\ -e_{2n}(z) &
\qquad z \in I_+
\end{array} \right. \qquad n\in\mathbb{N}. \]
This leads to
\[ \langle e_{-r}, V e_{2n} \rangle =
-\frac{\I (-1)^n}{\pi} \cos \left(
\frac{r\pi}{2} \right) \frac{1}{(2n+r)(2n+r+1)}, \]
so that
\[ \Sigma_3 = \frac{64}{\pi^4} \sum_{n=1}^\infty
\sum_{m=1}^\infty  \left( \sum_{l=0}^\infty
\sigma_{n,m,l}^{(3)} \right)^2 \]
where
\[ \sigma_{n,m,l}^{(3)} = \frac{(-1)^l}{(4n-2l+1)(4n-2l+3)
(4m+2l+1)(4m+2l+3)}. \]
Further,
\begin{eqnarray*}
\Sigma_4 &=& \| P_2 {V'}^* S_{\rm NS} V P_4 \|_2^2
= \sum_{n=1}^\infty \sum_{m=1}^\infty \left|
\sum_{r \in \mathbb{N}_0 +\frac{1}{2}}
\overline{ \langle e_{-r}, V' e_{2n+1} \rangle }
\langle e_{-r}, V e_{2m} \rangle \right|^2 \\
&=& \frac{64}{\pi^4} \sum_{n=1}^\infty \sum_{m=1}^\infty
\left( \sum_{l=0}^\infty \sigma_{n,m,l}^{(4)} \right)^2
\end{eqnarray*}
where
\[ \sigma_{n,m,l}^{(4)} = \frac{(-1)^l}{(4n+2l+3)(4n+2l+5)
(4m+2l+1)(4m+2l+3)}. \]
In the same way we compute
\[ \Sigma_5 = \| P_3 V^* S_{\rm NS} V P_4 \|_2^2
= \sum_{n=1}^\infty \sum_{m=1}^\infty \left|
\sum_{r \in \mathbb{N}_0 +\frac{1}{2}}
\overline{ \langle e_{-r}, V e_{-2n} \rangle }
\langle e_{-r}, V e_{2m} \rangle \right|^2 .\]
The action of $V$ on basis vectors $e_{-2n}$ is
\[ (Ve_{-2n})(z) = \left\{ \begin{array}{cc} e_{-2n}(z) &
\qquad z \in I_- \\
-\I e_{-2n-1}(z) & \qquad z \in I_2 \\ -e_{-2n}(z) & \qquad z
\in I_+ \end{array} \right. \qquad n\in\mathbb{N}.\]
This leads us to
\[ \langle e_{-r}, V  e_{-2n} \rangle =  \frac{\I (-1)^n}{\pi}
\cos\left(\frac{r\pi}{2}\right)\frac{1}{(2n-r)(2n-r+1)},\]
so that
\[ \Sigma_5 = \frac{64}{\pi^4} \sum_{n=1}^\infty
\sum_{m=1}^\infty  \left( \sum_{l=0}^\infty
\sigma_{n,m,l}^{(5)} \right)^2 \]
where
\[ \sigma_{n,m,l}^{(5)} = \frac{1}{(4n-2l-1)(4n-2l+1)
(4m+2l+1)(4m+2l+3)}. \]
Finally,
\begin{eqnarray*}
\Sigma_6 &=& \| P_4 V^* S_{\rm NS} V P_4 \|_2^2
= \sum_{n=1}^\infty \sum_{m=1}^\infty \left|
\sum_{r \in \mathbb{N}_0 +\frac{1}{2}}
\overline{ \langle e_{-r}, V e_{2n} \rangle }
\langle e_{-r}, V e_{2m} \rangle \right|^2 \\
&=& \frac{64}{\pi^4} \sum_{n=1}^\infty \sum_{m=1}^\infty
\left( \sum_{l=0}^\infty \sigma_{n,m,l}^{(6)} \right)^2
\end{eqnarray*}
where
\[ \sigma_{n,m,l}^{(6)} = \frac{1}{(4n+2l+1)(4n+2l+3)
(4m+2l+1)(4m+2l+3)}. \]
Next, we turn to the discussion of the operator
\[ C=VS_{\rm NS}V^*-S_{\rm R}. \]
For showing that $\|C\|_2<\infty$ we prove that
\[ \| ({\bf 1}-P_0)C({\bf 1}-P_0) \|_2 =
\left\| \sum_{i,j=1}^4 P_i C P_j \right\|_2 \le
\sum_{i,j=1}^4 \|P_i C P_j \|_2 < \infty. \]
Because $C=C^*$ we have again only to treat those terms with
$i\le j$. Further, by
\[ \Gamma C \Gamma = V({\bf 1}-S_{\rm NS})V^*-({\bf 1}-S_{\rm R})
= (VV^*-{\bf 1}) - C \]
and since ${\bf 1}-VV^*$ is a rank one projection
(i.e. $\|VV^*-{\bf 1}\|_2=1$), we find
\begin{eqnarray*}
\|P_1CP_1\|_2 &=& \| \Gamma P_1CP_1 \Gamma \|_2 =
\|P_2 \Gamma C \Gamma P_2 \|_2 =
\|P_2(VV^*-{\bf 1}-C)P_2\|_2 \\
&\le& \|P_2CP_2\|_2+1.
\end{eqnarray*}
In the same way one obtains
\[ \| P_3 C P_3 \|_2  \le \| P_4 C P_4 \|_2+1,\qquad
\| P_2 C P_3 \|_2 \le \| P_1 C P_4 \|_2+1, \]
and
\[ \| P_1 C P_3 \|_2 \le \| P_2 C P_4 \|_2+1. \]
Again, $S_{\rm R}$ is annihilated by $P_2$ or $P_4$ in
these terms. Using once more that $V-V'$ is
Hilbert Schmidt class, we conclude that it is
sufficient to prove the finiteness of the
following six terms:
\begin{eqnarray*}
\|P_2 V S_{\rm NS} V^* P_2 \|_2,\quad
\|P_1 V S_{\rm NS} V^* P_2 \|_2,\quad
\|P_1 V S_{\rm NS} {V'}^* P_4 \|_2,\\
\|P_2 V S_{\rm NS} {V'}^* P_4 \|_2,\quad
\|P_3 V' S_{\rm NS} {V'}^* P_4 \|_2,\quad
\|P_4 V' S_{\rm NS} {V'}^* P_4 \|_2.
\end{eqnarray*}
Now we have to work again,
\[ \Sigma_7 = \| P_2 V S_{\rm NS} V^* P_2 \|_2^2
= \sum_{n=1}^\infty \sum_{m=1}^\infty \left|
\sum_{r \in \mathbb{N}_0 +\frac{1}{2}}
\overline{ \langle e_{-r}, V^* e_{2n+1} \rangle }
\langle e_{-r}, V^* e_{2m+1} \rangle \right|^2 .\]
The action of $V^*$ on basis vectors $e_{2n+1}$ is
\[ (V^*e_{2n+1})(z) = \left\{ \begin{array}{cc} e_{2n+1}(z) &
\qquad z \in I_- \\
-\I e_{2n}(z) & \qquad z \in I_2 \\ -e_{2n+1}(z) & \qquad z
\in I_+ \end{array} \right. \qquad n\in\mathbb{N}.\]
This leads us to
\[ \langle e_{-r}, V^*  e_{2n+1} \rangle =  -\frac{\I (-1)^n}{\pi}
\sin\left(\frac{r\pi}{2}\right)\frac{1}{(2n+r)(2n+r+1)},\]
so that
\[ \Sigma_7 = \frac{64}{\pi^4} \sum_{n=1}^\infty
\sum_{m=1}^\infty  \left( \sum_{l=0}^\infty
\sigma_{n,m,l}^{(7)} \right)^2 \]
where
\[ \sigma_{n,m,l}^{(7)} = \frac{1}{(4n+2l+1)(4n+2l+3)
(4m+2l+1)(4m+2l+3)}. \]
Further,
\[ \Sigma_8 = \| P_1 V S_{\rm NS} V^* P_2 \|_2^2
= \sum_{n=1}^\infty \sum_{m=1}^\infty \left|
\sum_{r \in \mathbb{N}_0 +\frac{1}{2}}
\overline{ \langle e_{-r}, V^* e_{-2n-1} \rangle }
\langle e_{-r}, V^* e_{2m+1} \rangle \right|^2 .\]
The action of $V^*$ on basis vectors $e_{-2n-1}$ is
\[ (V^*e_{-2n-1})(z) = \left\{ \begin{array}{cc} e_{-2n-1}(z) &
\qquad z \in I_- \\
\I e_{-2n}(z) & \qquad z \in I_2 \\ -e_{-2n-1}(z) & \qquad z
\in I_+ \end{array} \right. \qquad n\in\mathbb{N}.\]
This leads us to
\[ \langle e_{-r}, V^*  e_{-2n-1} \rangle =  -\frac{\I (-1)^n}{\pi}
\sin\left(\frac{r\pi}{2}\right)\frac{1}{(2n-r)(2n-r+1)},\]
so that
\[ \Sigma_8 = \frac{64}{\pi^4} \sum_{n=1}^\infty
\sum_{m=1}^\infty  \left( \sum_{l=0}^\infty
\sigma_{n,m,l}^{(8)} \right)^2 \]
where
\[ \sigma_{n,m,l}^{(8)} = \frac{1}{(4n-2l-1)(4n-2l+1)
(4m+2l+1)(4m+2l+3)}. \]
Further,
\[ \Sigma_9 = \| P_1 V S_{\rm NS} {V'}^* P_4 \|_2^2
= \sum_{n=1}^\infty \sum_{m=1}^\infty \left|
\sum_{r \in \mathbb{N}_0 +\frac{1}{2}}
\overline{ \langle e_{-r}, V^* e_{-2n-1} \rangle }
\langle e_{-r}, {V'}^* e_{2m} \rangle \right|^2 .\]
The action of ${V'}^*$ on basis vectors $e_{2n}$ is
\[ ({V'}^*e_{2n})(z) = \left\{ \begin{array}{cc} e_{2n}(z) &
\qquad z \in I_- \\
-\I e_{2n-1}(z) & \qquad z \in I_2 \\ -e_{2n}(z) & \qquad z
\in I_+ \end{array} \right. \qquad n\in\mathbb{N}.\]
This leads us to
\[ \langle e_{-r}, {V'}^*  e_{2n} \rangle =  \frac{\I (-1)^n}{\pi}
\cos\left(\frac{r\pi}{2}\right)\frac{1}{(2n+r)(2n+r-1)},\]
so that
\[ \Sigma_9 = \frac{64}{\pi^4} \sum_{n=1}^\infty
\sum_{m=1}^\infty  \left( \sum_{l=0}^\infty
\sigma_{n,m,l}^{(9)} \right)^2 \]
where
\[ \sigma_{n,m,l}^{(9)} = \frac{(-1)^l}{(4n-2l-1)(4n-2l+1)
(4m+2l-1)(4m+2l+1)}. \]
Further,
\begin{eqnarray*}
\Sigma_{10} &=& \| P_2 V S_{\rm NS} {V'}^* P_4 \|_2^2
= \sum_{n=1}^\infty \sum_{m=1}^\infty \left|
\sum_{r \in \mathbb{N}_0 +\frac{1}{2}}
\overline{ \langle e_{-r}, V^* e_{2n+1} \rangle }
\langle e_{-r}, {V'}^* e_{2m} \rangle \right|^2 \\
&=& \frac{64}{\pi^4} \sum_{n=1}^\infty
\sum_{m=1}^\infty  \left( \sum_{l=0}^\infty
\sigma_{n,m,l}^{(10)} \right)^2
\end{eqnarray*}
where
\[ \sigma_{n,m,l}^{(10)} = \frac{(-1)^l}{(4n+2l+1)(4n+2l+3)
(4m+2l-1)(4m+2l+1)}. \]
Further,
\[ \Sigma_{11} = \| P_3 V' S_{\rm NS} {V'}^* P_4 \|_2^2
= \sum_{n=1}^\infty \sum_{m=1}^\infty \left|
\sum_{r \in \mathbb{N}_0 +\frac{1}{2}}
\overline{ \langle e_{-r}, {V'}^* e_{-2n} \rangle }
\langle e_{-r}, V e_{2m} \rangle \right|^2 .\]
The action of ${V'}^*$ on basis vectors $e_{-2n}$ is
\[ ({V'}^*e_{-2n})(z) = \left\{ \begin{array}{cc} e_{-2n}(z) &
\qquad z \in I_- \\
\I e_{-2n+1}(z) & \qquad z \in I_2 \\ -e_{-2n}(z) & \qquad z
\in I_+ \end{array} \right. \qquad n\in\mathbb{N}.\]
This leads us to
\[ \langle e_{-r}, {V'}^*  e_{-2n} \rangle =
-\frac{\I (-1)^n}{\pi}
\cos\left(\frac{r\pi}{2}\right)\frac{1}{(2n-r)(2n-r-1)},\]
so that
\[ \Sigma_{11} = \frac{64}{\pi^4} \sum_{n=1}^\infty
\sum_{m=1}^\infty  \left( \sum_{l=0}^\infty
\sigma_{n,m,l}^{(11)} \right)^2 \]
where
\[ \sigma_{n,m,l}^{(11)} = \frac{1}{(4n-2l-3)(4n-2l-1)
(4m+2l-1)(4m+2l+1)}. \]
Finally,
\begin{eqnarray*}
\Sigma_{12} &=& \| P_4 V' S_{\rm NS} {V'}^* P_4 \|_2^2
= \sum_{n=1}^\infty \sum_{m=1}^\infty \left|
\sum_{r \in \mathbb{N}_0 +\frac{1}{2}}
\overline{ \langle e_{-r}, {V'}^* e_{2n} \rangle }
\langle e_{-r}, {V'}^* e_{2m} \rangle \right|^2 \\
&=& \frac{64}{\pi^4} \sum_{n=1}^\infty
\sum_{m=1}^\infty  \left( \sum_{l=0}^\infty
\sigma_{n,m,l}^{(12)} \right)^2
\end{eqnarray*}
where
\[ \sigma_{n,m,l}^{(12)} = \frac{1}{(4n+2l-1)(4n+2l+1)
(4m+2l-1)(4m+2l+1)}. \]
We have the following estimate of absolute values
of the $\sigma_{n,m,l}^{(j)}$ for $j=1,4,6,7,$ $10,12$:
\begin{equation}
\label{est}
\sigma_{n,m,l}^{(1)} \ge |\sigma_{n,m,l+2}^{(j)}|,
\qquad j=4,6,7,10,12, \quad n,m\in\mathbb{N},\quad
l\in\mathbb{N}_0.
\end{equation}
If we omit in our summations $l=0$ and $l=1$ terms,
this corresponds to the replacement of $S_{\rm NS}$ by
\[ S_{\rm NS}' = S_{\rm NS} - |e_{-\frac{1}{2}}\rangle\langle
e_{-\frac{1}{2}}| - |e_{-\frac{3}{2}}\rangle\langle
e_{-\frac{3}{2}}|. \]
Since the difference $S_{\rm NS}-S_{\rm NS}'$ is obviously
Hilbert Schmidt class, this has no influence of
the property of $\Sigma_j$ to be finite or infinite.
Hence the estimate (\ref{est}) tells us that for
the proof of $\Sigma_j<\infty$, $j=1,4,6,7,10,12$,
it is sufficient to prove it for $j=1$. We compute
\begin{eqnarray*}
\lefteqn{\Sigma_1=}\\
&=& \frac{64}{\pi^4} \sum_{n,m=1}^\infty
\left( \sum_{l=0}^\infty
\frac{1}{(4n+2l+3)(4n+2l+5)(4m+2l+3)(4m+2l+5)}
\right)^2 \\
&<& \frac{64}{\pi^4} \sum_{n=1}^\infty
\sum_{m=1}^\infty \frac{1}{(2m+1)^4}
\left( \sum_{l=0}^\infty \frac{1}{(4n+2l+3)^2}
\right)^2 \\
&=& \frac{2}{3} \sum_{n=1}^\infty \left(
\sum_{l=0}^\infty \frac{1}{(4n+2l+3)^2}
\right)^2 \\
&\le& \frac{2}{3} \sum_{n=1}^\infty \left(
\int_0^\infty \frac{dl}{(4n+2l+1)^2} \right)^2 \\
&=& \frac{2}{3} \sum_{n=1}^\infty \left(
\frac{1}{2(4n+1)} \right)^2 \\
&<& \frac{1}{96} \sum_{n=1}^\infty \frac{1}{n^2} \\
&=& \frac{\pi^2}{576},
\end{eqnarray*}
$\Sigma_1$ is finite. On the other hand we find
for $n,m\in\mathbb{N}$, $l\in\mathbb{N}_0$
\begin{eqnarray*}
|\sigma_{n,m,l}^{(11)}| &>& |\sigma_{n,m,l+2}^{(2)}|,\qquad
|\sigma_{n,m,l}^{(11)}| > |\sigma_{n,m,l+2}^{(3)}|,\qquad
|\sigma_{n,m,l}^{(11)}| > |\sigma_{n,m,l+1}^{(5)}|,\\
|\sigma_{n,m,l}^{(11)}| &>& |\sigma_{n,m,l+1}^{(8)}|,\qquad
|\sigma_{n,m,l}^{(11)}| > |\sigma_{n,m,l+1}^{(9)}|.
\end{eqnarray*}
By the same argument, for the proof of
$\Sigma_j<\infty$, $j=2,3,5,8,9,11$, it is
sufficient to prove that
\[ \tilde{\Sigma}_{11} = \frac{64}{\pi^4} \sum_{n=1}^\infty
\sum_{m=1}^\infty \left( \sum_{l=0}^\infty
|\sigma_{n,m,l}^{(11)}| \right)^2 < \infty. \]
For this purpose, we decompose the sum over the index
$l$ into three parts,
\begin{eqnarray*}
\sum_{l=0}^\infty |\sigma_{n,m,l}^{(11)}| &=&
\sum_{l=0}^{2n-2} |\sigma_{n,m,l}^{(11)}| +
|\sigma_{n,m,2n-1}^{(11)}| + \sum_{l=2n}^\infty
|\sigma_{n,m,l}^{(11)}| \\ &=&
\sum_{l=0}^{2n-2} \sigma_{n,m,l}^{(11)} -
\sigma_{n,m,2n-1}^{(11)} + \sum_{l=2n}^\infty
\sigma_{n,m,l}^{(11)}.
\end{eqnarray*}
We begin with estimating the first part. By reversing
the order of summation we obtain
\begin{eqnarray*}
\lefteqn{\sum_{l=0}^{2n-2} \sigma_{n,m,l}^{(11)}=}\\
&=& \sum_{l=0}^{2n-2}
\frac{1}{(2l+1)(2l+3)(4m+4n-2l-3)(4m+4n-2l-5)} \\
&<& \sum_{l=0}^{2n+2m-3}
\frac{1}{(2l+1)^2(4n+4m-2l-5)^2} \\
&=& \sum_{l=0}^{n+m-2}\frac{1}{(2l+1)^2(4n+4m-2l-5)^2}
\\ && \qquad + \sum_{l=n+m-1}^{2n+2m-3}
\frac{1}{(2l+1)^2(4n+4m-2l-5)^2} \\
&=& 2\sum_{l=0}^{n+m-2}\frac{1}{(2l+1)^2(4n+4m-2l-5)^2}\\
&\le& \frac{2}{(4n+4m-5)^2} + 2 \int_0^{n+m-2}
\frac{dl}{(2l+1)^2(4n+4m-2l-5)^2} \\
&=& \frac{2}{(4n+4m-5)^2} +  \int_1^{2n+2m-3}
\frac{dx}{x^2(4n+4m-4-x)^2} \\
&=& \frac{2}{(4n+4m-5)^2} + \frac{1}{(4n+4m-4)^2(2n+2m-1)}
\\ && \quad -\frac{1}{(4n+4m-4)^2(4n+4m-5)}  \\
&& \quad
-\frac{1}{(4n+4m-4)(2n+2m-3)(2n+2m-1)} \\
&& \quad
+\frac{1}{(4n+4m-4)(4n+4m-5)}
-\frac{2}{(4n+4m-4)^3}
\ln\left(\frac{2n+2m-1}{2n+2m-3}\right)
\\ && \qquad
+\frac{2}{(4n+4m-4)^3} \ln ( 4n+4m-5) \\
&<& \frac{6}{(2n+2m-3)^2}.
\end{eqnarray*}
The last estimate is very rough but correct. In our computation
we have used the fact that in an area of strict decrease a
summation can be estimated by an integral plus the first
summand. Next we consider the only negative term,
\[ -\sigma_{n,m,2n-1}^{(11)} = \frac{1}{(4n+4m-3)(4n+4m-1)}
< \frac{1}{(2n+2m-3)^2}, \]
and finally the remaining summation,
\begin{eqnarray*}
\sum_{l=2n}^\infty \sigma_{n,m,l}^{(11)}
&=&\sum_{l=0}^\infty
\frac{1}{(2l+1)(2l+3)(4n+4m+2l-1)(4n+4m+2l+1)} \\
&<& \frac{1}{(4n+4m-1)^2} \sum_{l=0}^\infty
\frac{1}{(2l+1)(2l+3)} \\
&=& \frac{1}{(4n+4m-1)^2} \\
&<& \frac{1}{(2n+2m-3)^2}.
\end{eqnarray*}
We now can conclude that
\[ \tilde{\Sigma}_{11} < \frac{64}{\pi^4} \sum_{n=1}^\infty
\sum_{m=1}^\infty \left( \frac{8}{(2n+2m-3)^2} \right)^2
= \frac{4096}{\pi^4} \sum_{k=0}^\infty
\frac{k+1}{(2k+1)^4} < \infty, \]
the proof of Lemma \ref{HS} is complete, q.e.d.
\end{appendix}

{\it acknowledgement}
I am grateful to Prof.~K. Fredenhagen
for discussions. He supported this investigation with
many ideas. Thanks are also due to C. Binnenhei
for a careful reading of the manuscript.


\begin{thebibliography}{99}
\bibitem{Ara1} Araki, H.: On Quasifree States
of CAR and Bogoliubov Automorphisms. Publ.~RIMS Kyoto
Univ.~Vol.~{\bf 6} (1970/71)
\bibitem{Ara2} Araki, H.: Bogoliubov Automorphisms
and Fock Representations of the Canonical Anticommutation
Relations. In: Operator Algebras and Mathematical
Physics, Am. Math.~Soc.~Vol.~{\bf 62} (1987)
\bibitem{AE} Araki, H., Evans, D.E.: On a
$C^*$-Algebra Approach to Phase Transitions in the
Two-Dimensional Ising Model.
Commun.~Math.~Phys.~{\bf 91} (1983)
\bibitem{Binnenneu} Binnenhei, C.: Implementation of
Endomorphisms of the CAR algebra. SFB 288 preprint
No.~142. Submitted to Rev.~Math.~Phys. (1994)
\bibitem{ich}  B{\"o}ckenhauer, J.: Lokale
Normalit{\"a}t und lokalisierte Endomorphismen des chiralen
Ising-Modells. Diplomarbeit, Hamburg (1994)
\bibitem{Decom} B{\"o}ckenhauer, J.: Decomposition
of Representations of CAR Induced by Bogoliubov
Endomorphisms. DESY 94-173 (1994)
\bibitem{BGL} Brunetti, R., Guido, D., Longo, R.:
Modular Structure and Duality in Conformal Quantum
Field Theory. Commun.~Math.~Phys.~{\bf 156} (1993)
\bibitem{BMT} Buchholz, D., Mack, G., Todorov, I.:
Localized Automorphisms of the U(1)-Current
Algebra on the Circle: An Instructive Example.
In: \cite{Kast}
\bibitem{Buch} Buchholz, D., Schulz-Mirbach, H.: Haag
Duality in Conformal Quantum Field Theory.
Rev.~Math.~Phys.~{\bf 2} (1990)
\bibitem{DHR1} Doplicher, S., Haag, R., Roberts, J.E.:
Fields, Observables and Gauge Transformations I \& II.
Commun.~Math.~Phys.~{\bf 13} (1969) and {\bf 15} (1969)
\bibitem{DHR2} Doplicher, S., Haag, R., Roberts, J.E.:
Local Observables and Particle Statistics I \&  II.
Commun.~Math.~Phys.~{\bf 23} (1971) and {\bf 35} (1974)
\bibitem{Doug} Douglas, R.G.: Banach Algebra
Techniques in Operator Theory. Academic Press, New York,
London (1972)
\bibitem{FRS1} Fredenhagen, K., Rehren, K.-H., Schroer, B.:
Superselection Sectors with Braid Group Statistics and
Exchange Algebras I. Commun.~Math. Phys.~{\bf 125} (1989)
\bibitem{FRS2} Fredenhagen, K., Rehren, K.-H., Schroer, B.:
Superselection Sectors with Braid Group Statistics and
Exchange Algebras II. SFB 288 preprint No.~10 (1992)
\bibitem{Fred3} Fredenhagen, K.: Quantum
Field Theory on Nontrivial Spacetimes. To appear in:
Proceedings of the Beersheva Conference 1993, R. Sen (Hrsg.)
\bibitem{FGV} Fuchs, J., Ganchev, A., Vecserny{\'e}s, P.:
Level 1 WZW Superselection Sectors.
Commun. Math.~Phys.~{\bf 146} (1992)
\bibitem{Haag} Haag, R.: Local Quantum Physics.
Springer-Verlag, Berlin, Heidelberg, New York (1992)
\bibitem{HK} Haag, R., Kastler, D.: An Algebraic Approach
to Quantum Field Theory. J.~Math.~Phys. {\bf 5} (1964)
\bibitem{Kast} Kastler, D. (ed.):  The Algebraic
Theory of Superselection Sectors. World Scientific,
Singapur (1990)
\bibitem{Lang} Lang, S.: ${\rm SL}_2(\mathbb{R})$.
Springer-Verlag, Berlin, Heidelberg, New York (1985)
\bibitem{Loke} Loke, T.: Operator Algebras and
Conformal Field Theory of the Discrete Series
Representations of Diff$(S^1)$. Dissertation,
Cambridge (1994)
\bibitem{Mack} L{\"u}scher, M., Mack, G.:  The
Energy Momentum Tensor of Critical Quantum Field Theories
in 1+1 Dimensions. Unpublished manuscript (1976)
\bibitem{MS1} Mack, G., Schomerus, V.: Conformal
Field Algebras with Quantum Symmetry from the Theory of
Superselection Sectors. Commun.~Math. Phys.~{\bf 134} (1990)
\bibitem{MS2} Mack, G., Schomerus, V.:
Endomorphisms and Quantum Symmetry of the Conformal
Ising Model. In: \cite{Kast}
\bibitem{Pow} Powers, R.T., St{\o}rmer, E.: Free
States of the Canonical Anticommutation Relations.
Commun.~Math.~Phys.~{\bf 16} (1970)
\bibitem{RS1} Reed, M., Simon, B.: Methods of Modern
Mathematical Physics. Vol.~{\bf 1}: Functional
Analysis. Academic Press, New York, London (1972)
\bibitem{Rideau} Rideau, G.: On Some Representations
of the Anticommutation Relations.
Commun. Math.~Phys.~{\bf 9} (1968)
\bibitem{Berti} Schroer, B.: A Trip to
Scalingland. In: Ferreira, E. (ed.): V. Brazilian
Symposium on Theoretical Physics, Vol. I (1974)
\bibitem{Szl1} Szlach\'{a}nyi, K.: Chiral
Decomposition as a Source of Quantum Symmetry in the
Ising Model. KFKI-1993-16/A preprint (1993)
\bibitem{Szl2} Szlach\'{a}nyi, K.: The Universal Algebra
of Local Even CAR and Majorana Algebras on the Circle.
Unpublished manuscript (1993)
\bibitem{Wass} Wassermann, A.: Operator Algebras and
Conformal Field Theory. To appear in: Proceedings of
the International Congress of Mathematicians,
Z{\"u}rich 1994, Birkh{\"a}user Verlag
\end{thebibliography}
\end{document}